\documentclass[11pt]{article}

\usepackage{amsmath,mathrsfs,amsbsy,amsthm}
\usepackage{cite}
\usepackage{graphicx,epsfig}
\usepackage{latexsym,amssymb}
\usepackage{epsf}
\usepackage{ifpdf,lineno}
\usepackage{color}
\usepackage{geometry}
\usepackage{amsfonts}
\usepackage{booktabs}
\usepackage{siunitx}
\usepackage{graphicx,subcaption}
\usepackage[export]{adjustbox}
\usepackage{tikz}

\def\sech{\hspace{0.75mm}{\rm sech}}
\def\sech{\hspace{0.75mm}{\rm sech}}
	
\begin{document}
	
	\title{{\bf Scattering of kinks in the $B\varphi^{4}$ model}}

\author{M. Mohammadi$^{1}$\thanks{Corresponding Author. } \\
	{\small \texttt{physmohammadi@pgu.ac.ir}}\and
	E. Momeni$^{2}$ \\{\small \texttt{ehsan.momeni@studenti.unipd.it}}}
	\date{{\em{$^1$Physics Department, Persian Gulf University, Bushehr 75169, Iran.\\
$^2$  Department of Physics and Astronomy, University of Padova, Via Marzolo, 8 - 35131 Padova, Italy.}}}
	\maketitle

\begin{abstract}

In this study, based on the $\varphi^4$ model, a new model (called the $B\varphi^4$ model) is introduced in which the potential form for the values of the field whose magnitudes are greater than $1$ is multiplied by the positive number $B$.
All features related to a single kink (antikink) solution remain unchanged and are independent of parameter $B$. However, when a kink interacts with an antikink in a collision, the results will significantly depend on parameter $B$.
Hence, for kink-antikink collisions, many features such as the critical speed, output velocities for a fixed initial speed, two-bounce escape windows, extreme values, and fractal structure in terms of parameter $B$ are considered in detail numerically.
The role of parameter $B$ in the emergence of a nearly soliton behavior in kink-antikink collisions at some initial speed intervals is clearly confirmed. The fractal structure in the diagrams of escape windows is seen for the regime $B\leq 1$.
However, for the regime $B >1$, this behavior gradually becomes fuzzing and chaotic as it approaches $B = 3.3$. The case $B = 3.3$ is obtained again as the minimum of the critical speed curve as a function of $B$. For the regime $3.3< B \leq 10$, the chaotic behavior gradually decreases. However, a fractal structure is never observed. Nevertheless, it is shown that despite the fuzzing and shuffling of the escape windows, they follow the rules of the resonant energy exchange theory.


\textbf{Keywords} : {Soliton, Kink Scattering, Resonance Phenomena, Topological Defect, Fractal. \\}

\end{abstract}


	

	\section{Introduction}\label{sec1}

Over the last few decades, nonlinear wave equations with solitary wave or especially soliton
solutions have played a significant role in describing various phenomena in
different branches of physics including optics \cite{optic1,optic2,optic3,optic4}, condensed matter \cite{condense1,condense2,condense3}, high energy physics \cite{HEP2,HEP3},
biophysics \cite{bio}, and so on.
Solitary wave solutions (also called defect structures) are the special
solutions of nonlinear wave equations that can freely propagate without any distortion in their profiles. In addition, their corresponding energy densities are localized.
Solitons are a  special type of solitary wave solutions whose profiles and velocities are restored without any change after collisions. In general, solitary wave solutions can be classified into classical or relativistic and topological or non-topological.
The topological property is important because by considering it the solitary wave solution will be inherently stable and minimally energized.
As examples of the relativistic topological
defects in $3+1$ dimensions, one can mention strings, vortices, magnetic monopoles, and
skyrmions \cite{HEP2,HEP3,SKrme,toft,MKP,str1,vor1} which are applied in cosmology as well as hadron and nuclear
physics.
Q-balls can be mentioned in relation to relativistic non-topological solutions \cite{Qball1,Qball2,Qball3}.  Moreover, there are a great number of non-relativistic defect structures (topological or non-topological) among which one can name the soliton solutions of the KdV, nonlinear
Schr\"{o}dinger, and Burgers equations \cite{nonrela1,nonrela2,nonrela3}.

The kink (antikink) solutions of the nonlinear Klein-Gordon equations in $1+1$ dimensions are the simplest relativistic topological defect structures \cite{HEP2,HEP3}.
The importance of such solutions has been explicitly demonstrated in studying wave motion in DNA molecules and graphene sheets  \cite{bio,DNA1,DNA2,DNA3,graphine0,graphine1,graphine2,graphine3,graphine4}.
Furthermore, in cosmology, the structure and dynamics of domain walls in $3+1$ dimensions can be correctly described by $1 + 1$-dimensional kink-bearing theories \cite{cosmology1,cosmology2,cosmology3,cosmology4,cosmology5,cosmology6,cosmology7,cosmology8,cosmology9}.
Among the great variety of systems with kink (antikink) solutions, the two systems of $\varphi^4$  and sine-Gordon (SG) have been particularly studied \cite{HEP2,HEP3}.
The sine-Gordon system is important because it is the only integrable Klein-Gordon kink-bearing system which yields soliton solutions.
The double SG (DSG) model which is a modified version of the SG model has also received much attention \cite{DSG1,DSG2,DSG3,DSG5}.
Interestingly, the DSG system is used for the description of some physical systems such as gold dislocations \cite{ADSG1}, optical pulses \cite{ADSG2}, and Josephson structures \cite{ADSG3}.

The $\varphi^4$ system and its deformed versions have been widely studied due to their simplicity
and applications\cite{graphine1,phi48,phi49,phi410,phi411,phi412,phi4150,phi415,phi4152,phi4160,phi416,phi417,phi419,phi420,phi44,phi46,phi477,phi4csf,phi47,phi4cc,phi4cc2,book}.
In relation to the study of kink-antikink scattering and the quasi-fractal structure, the $\varphi^4$ system has been considered as a representative of such non-integrable systems \cite{phi44,phi46,phi477,phi4csf,phi47,phi4cc,phi4cc2}.
It has been studied from different aspects including the interaction of kinks with impurities \cite{phi48,phi49,phi410,phi411}, maximum values of quantities  in a kink-antikink collision \cite{phi412}, scattering between wobbling kink and antikink \cite{phi4150,phi415,phi4152},  the standard deformed versions \cite{phi4160,phi416,phi417,phi419,phi420}, and  collective coordinate models for kink-antikink interactions \cite{phi47,phi4cc,phi4cc2}. There is also an informative book on the subject that specifically examines the $\varphi^4$ system from various angles \cite{book}. 
Recently, models with polynomial potentials of higher degrees have also received a lot of attention including the $\varphi^6$ model \cite{phi61,phi62,phi63}, the $\varphi^8$ model \cite{phi81,Gani.PRD.2020, Gani.EPJC.2021,phi82}, and the models that yield kinks with power-law tails \cite{phi83,polonomial1,polonomial2,polonomial3,polonomial4,polonomial5}. Two scalar field models have also been of interest to researchers \cite{cosmology2,cosmology3,cosmology4,couple1,couple2,couple3,couple4,couple5,couple6,couple7}.

Kink-antikink collisions exhibit a richer behavior in non-integrable systems (such as the $\varphi^4$ system) than in integrable ones \cite{phi44,phi46,phi47,phi477}.
In this respect, except for the SG and radiative systems, the study of output velocity in terms of the incoming (initial) velocity for non-integrable kink-bearing systems has usually led to the similar results.
First, kink-antikink collisions face the two different fates of scattering or bion formation. Bion is a metastable non-topological object which decays slowly and radiates energy in the form of small-amplitude waves.
Second, there is always a critical velocity ($v_{cr}$) after which a bionic state can no longer be formed. Third, for incoming velocities less than the critical velocity, the typical fate is the formation of a bionic state. However, there are many wide and narrow intervals of initial velocities that lead to escape, are labelled with integer numbers ${\cal N}>1$, and are called ${\cal N}$-bounce escape  windows.
More precisely, an ${\cal N}$-bounce escape  window is an interval of incoming velocities in which the kink and antikink collide ${\cal N}$ times before escaping. In such a regime of initial velocities (i.e. $v<v_{cr}$), a quasi-fractal structure is created from the arrangement of different escape  windows.

\begin{figure}[h!]
	\centering
	\includegraphics[scale=.6]{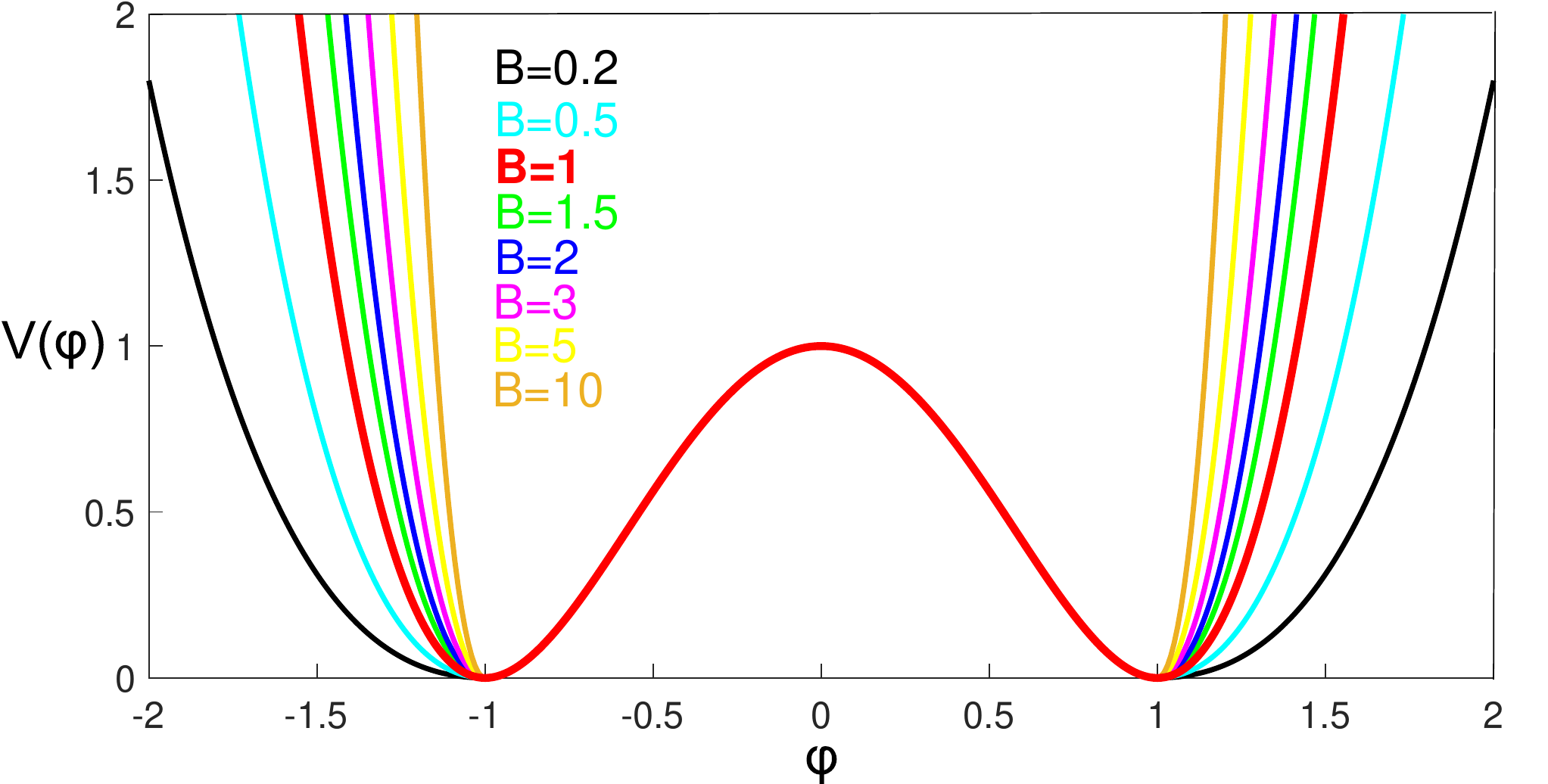}
	\caption{The red curve represents the potential of the $\varphi^4$ model. For  $B\varphi^4$ systems, the potential shape in the range of $-1$ to $1$ remains unchanged and is the same as that of the $\varphi^4$ model. Otherwise, it is multiplied by $B$. }\label{Pot}
\end{figure}

In this paper, the potential of the $\varphi^4$ system is deformed in a new and different way using parameter $B$ ($0.2\leq B \leq 10$) so that no change is made to the kink solution itself and the effect of this parameter can only be seen in the interactions.
In fact, the potential of the new system (called the $B\varphi^4$ system) can be divided into two parts. Part I is located between the two vacua $\varphi=-1$ and $\varphi=1$ and is the same as the $\varphi^4$ potential so that only this part of the potential is important for determining the kink (antikink) solution.
Part II (the rest) is the $\varphi^4$ potential multiplied by parameter $B$ at $\varphi<-1$ and $\varphi>1$ (see Fig.~\ref{Pot}). Indeed, studying different quantities related to kink-antikink collisions in terms of parameter $B$ is interesting and can enhance our knowledge of systems with kink solutions.
Also, from the physical point of view, the existence of systems with different potentials only over the kink sector is equivalent to the existence of the same structure (such as a DNA molecule, a graphene sheet or a domain wall)  but with different results in  interactions. For example, if  kink and antikink are considered as   two hypothetical classical  particles in $1+1$ dimensions,  different values of parameter $B$ lead to  different interactions for the same  particles, a situation which has no equivalent in  particle physics and can be interesting from this point of view.


Our study is structured as follows: In the next section, the $B\varphi^4$ model is introduced.
In addition, some necessary details and general properties of kink-bearing systems are briefly reintroduced.
In section 3, the necessary numerical considerations for the obtained results are presented. Section 4 deals with all numerical results obtained for many systems in the range of $0.2\leq B\leq 10$. 
This section contains six subsections in which several features related to kink-antikink collisions are studied in detail. In particular, the critical velocity, the escape  windows, the fractal structure, the two-bounce escape  windows, the output velocity for a fixed initial velocity, and the maximum values at the center of mass point are all studied in detail in terms of parameter $B$ in this section.
The last section is devoted to summary and conclusions. It should be noted that a lot of information is presented graphically in the form of high-resolution figures and can only be accessed by zooming in on them. Therefore, using the printed version of this article is not recommended at all.





\section{The $B\varphi^4$ model, the internal modes, and the conserved quantities}\label{sec2}

The real non-linear Klein-Gordon systems in $1+1$ dimensions can be generally introduced by the following action:
\begin{equation}\label{fv}
S=\int dx dt \left[ \frac{1}{2}\left(\frac{\partial\varphi}{\partial t}\right)^{2}-\frac{1}{2}\left(\frac{\partial\varphi}{\partial x}\right)^2-V(\varphi)\right] ,
 \end{equation}
where $\varphi$  is a real scalar field and $V(\varphi)$ represents a self-interaction term that is called the potential. In addition, we consider the velocity of light equal to one ($c=1$) throughout the present study. Using the least action principle, the dynamical equation for the evolution of $\varphi$ is obtained as follows:
\begin{equation}\label{dy}
 \ddot{\varphi}-\varphi''=-\frac{dV}{d\varphi},
 \end{equation}
where the primes and dots denote space and time derivatives, respectively. To have at
least one type of the well-known kink (anti-kink) solutions, the potential between two consecutive degenerate minima (vacua) must be positive.  The vacua are special values of the field  ($\varphi_{i}$)  for which both $V(\varphi_{i})$ and $\frac{dV}{d\varphi}|_{\varphi=\varphi_{i}}$  vanish. For instance, the potential of the  $\varphi^4$  model is
\begin{equation}\label{df}
	V(\varphi)=\dfrac{1}{2}(\varphi^2-1)^2,
\end{equation}
which has two  vacua  at   $\varphi_{1}=-1$ and $\varphi_{2}=1$ (see Fig.~\ref{Pot}).

Since the dynamical equation (\ref{dy}) is Lorentz invariant, by finding a non-moving
solution $\varphi_{o}(x,t)$, the moving version can be easily obtained just by applying a relativistic
boost.
In other words, one must replace $x$ and $t$ with $\bar{x}=\gamma(x-vt)$ and $\bar{t}=\gamma(t-vx)$, respectively. That is, the moving solution will be $\varphi_{v}(x,t)=\varphi_{o}(\bar{x},\bar{t})$, where $\gamma=1/\sqrt{1-v^2}$ and $v$ is any arbitrary velocity ($|v|<1$).
Based on this, the existence of a non-oscillating traveling wave solution ($\varphi_{v}(x,t)=\varphi_{o}(\bar{x})$) can be considered by examining the dynamical equation (\ref{dy}) for a static solution in the form of $\varphi_{o}(x)$. For example, for the $\varphi^4$ system  (\ref{df}), the special solution $\varphi_{o}(x)$ is obtained as:
\begin{equation}\label{ws}
\varphi_{o}(x)=\pm\tanh((x-x_{o})).
\end{equation}
where $x_{o}$ (the initial position) is any arbitrary constant. The moving version of this is as follows:
\begin{equation}\label{yh}
	\varphi_{v}(x,t)=\pm\tanh(\gamma(x-x_{o}-vt)),
\end{equation}
 In the above equations, the plus
(minus) sign is a kink solution that begins at $x=-\infty$ with $\varphi=-1$ ($\varphi=1$) and tends to
$\varphi=1$ ($\varphi=-1$) at $x=\infty$.

Hence, these topological solutions (\ref{yh}) vary between $\varphi_{1}=-1$ and $\varphi_{2}=1$ which is an interval between two consecutive vacua. In other words, it is only the shape of the potential between two consecutive vacua that determines the type of the kink and antikink solutions which do not depend on the potential form in other regions ($\varphi> 1$ and  $\varphi<-1$). In this respect, we can introduce the $B\varphi^4$ system as follows (see Fig.~\ref{Pot}):
\begin{equation}\label{cof}
	V(\varphi)=
	\begin{cases}
		\frac{1}{2}(\varphi^2-1)^2, & |\varphi|\leq 1,
		\\
		\frac{1}{2}B(\varphi^2-1)^2, & |\varphi|>1,
	\end{cases}
\end{equation}
where $B$ can be any arbitrary positive number. Therefore, since the $B\varphi^4$ systems have the same potential form in the range of $-1$ to $1$, they all have the same kink and anti-kink solutions (\ref{yh}) which are independent of parameter $B$.  It should be noted that at $\varphi=\pm 1$, the first derivative of potential (\ref{cof}) is continuous, but its second derivative is discontinuous. Also, the case $B=1$ is the ordinary $\varphi^4$ system (\ref{df}).


In general, for kink-bearing systems (\ref{fv}), there may be disturbed kink solutions which can be considered as a two-part superposition (a wobbling kink): a kink solution as a background and a low-amplitude oscillating function. More precisely, if $\varphi_{o}(x)$ is a static kink solution of Eq.~(\ref{dy}), its disturbed version will be:
\begin{equation}\label{fg}
	\varphi_{o}(x,t)=\varphi_{o}(x)+\psi(x)\sin(\omega_{o}t+\theta),
\end{equation}
where $\psi(x)$ is a small perturbation  ($|\psi|\ll 1$), $\theta$ is any arbitrary initial phase,
and $\omega_{o}$ is the rest frequency. By inserting Eq.~(\ref{fg}) into the nonlinear dynamical equation (\ref{dy}) and expanding it to the first order in  $\psi(x)$, an eigenvalue Schr\"{o}dinger-like equation is obtained for $\psi(x)$:
\begin{equation}\label{vb}
	-\psi''+V_{sch}(x)\psi=\omega_{o}^2\psi,
\end{equation}
where
\begin{equation}\label{fh}
	V_{sch}(x)=\left.\dfrac{d^2V}{d\varphi^2}\right|_{\varphi=\varphi_{o}},
\end{equation}
which is called the “\textit{schr\"{o}dinger potential}’’, the “\textit{kink potential}’’, or the “\textit{stability potential}’’.
Similarly, if the second derivative of the potential is continuous at a vacuum,  one can consider the low-amplitude perturbation around that. Thus, the Schr\"{o}dinger-like equation (\ref{vb}) is again obtained, except that $V_{sch}=\frac{d^2V}{d\varphi^2}|_{\varphi=\varphi_{i}}$ will now be a constant, and    $\varphi_{o}$ must be replaced with $\varphi_{i}$ and $\omega_{o}$ with $\omega_{i}$ in Eq.~(\ref{fg}).  
Quantum-mechanically speaking,   $\omega_{i}=(\frac{d^2V}{d\varphi^2}|_{\varphi=\varphi_{i}})^{1/2}$   corresponds to the mass threshold   of field quanta  around the vacuum $\varphi=\varphi_{i}$. In general, for a special sector $(\varphi_{i},\varphi_{i+1})$ corresponding to a special kink (antikink) solution, $\omega_{i}$ and  $\omega_{i+1}$ are not necessarily the same. In other words, $V_{sch}$  may have a jump at successive vacua, which is the case for the  $\varphi^6$ system \cite{phi61} but not for the $\varphi^4$ system.
For the next application, let us define the threshold frequency as $\omega_{th}=\min(\omega_{i},\omega_{i+1})$.  
In particular, for the  $\varphi^4$ system $\omega_{th}=2$,  
but for $B\varphi^4$ systems ($B\neq 1$), since the second derivative of the potential is discontinuous at vacua, it is impossible to define a  threshold frequency as usual. More precisely,   considering the low amplitude perturbations around vacua  cannot be introduced in a well-formed solution like (\ref{fg}) for $B\neq 1$, and Eq.~(\ref{vb}) is not valid for such perturbations.

The eigenvalue equation (\ref{vb}) always has a trivial solution $\psi\propto\frac{d\varphi_{o}}{dx}$ with $\omega_{o}=0$ which
corresponds only to an infinitesimal spatial translation of the static kink (antikink)
solution \cite{DSG1,phi47,AP}.
However, for some kink-bearing systems (\ref{fv}), there are one or more non-trivial solutions $\psi(x)$ with specific reset frequencies ($\omega_{o}$) called “\textit{internal modes}’’ or “\textit{shape modes}’’.
Kinks (antikinks) in systems with no non-trivial internal modes such as the SG system
cannot appear with a permanent oscillation after collisions.
In contrast to internal modes  as the discrete solutions of the eigenvalue equation (\ref{vb}) with the condition  $|\omega_{o}|<\omega_{th}$, the free modes are defined as the continuous solutions for $|\omega_{o}|>\omega_{th}$.
In general, internal modes (free modes) exponentially (periodically) tend to zero (oscillates as $\sin((\omega_{o}^2-\omega_{th}^2)^{1/2 }x
+\theta)$)
at large distances.
If $\omega_{th}=0$, we will have other types of systems called
“\textit{radiative systems}’’ which have their own characteristics \cite{RD,RD1}.
For $B\varphi^4$ systems,
 the kink potential is $V_{sch}(x)=4-6\sech^2(x)$, and Eq.~(\ref{vb}) leads to a non-trivial
 bound state (internal mode) corresponding to $\omega_{o}^2=3$ with  $\psi\propto \tanh(x)\sech(x)$, that is, the  same ones obtained for the $\varphi^4$  system \cite{DSG1,phi47}.
In relation to the free modes around a kink solution (not a vacuum), although a threshold frequency cannot be defined for $B\neq 1$, a similar situation exists with the $\varphi^4$ system. 
It should be noted that the moving version of the
static disturbed kink solution (\ref{fg}) can be obtained easily by replacing $t$ with $\gamma(t-vx)$ and
$x$ with $ \gamma (x-vt)$.


In examining the evolution of physical systems, it is important to find quantities that are conserved. First, examining these conserved quantities at different times can be used as the indicators of the validity and accuracy of our numerical method.
Second, some possible physical relations between these conserved quantities provide the basis for obtaining some other unknown quantities numerically. The next paragraph will give an example of the relations used to obtain the velocity of each entity. The conserved quantities for the kink-bearing systems (\ref{fv}) are topological charge (\ref{tc}), energy (\ref{en}), and momentum (\ref{bng}):
\begin{eqnarray} \label{tc}
	&& Q=\varphi(+\infty)-\varphi(-\infty),
	\\ \label{en}&& E[\varphi]=\int_{-\infty}^{+\infty}\varepsilon(x,t) ~dx=\int_{-\infty}^{+\infty}\left(\frac{1}{2}\dot{\varphi}^{2}+\frac{1}{2}\varphi'^{2} +V(\varphi)\right)dx=K+U+P,\quad\quad\quad
	\\ \label{bng}&& P[\varphi]=\int_{-\infty}^{+\infty}(-\dot{\varphi}\varphi') dx.
\end{eqnarray}
In Eq.~(\ref{en}), the total energy of the system ($E$) is written as the sum of three parts, i.e. the kinetic energy ($K$), the gradient energy ($U$), and the potential energy ($P$) which are defined as the integrations of $k(x,t)=\frac{1}{2}\dot{\varphi}^2$, $u(x,t)=\frac{1}{2}\varphi'^2 $, and $p(x,t)=V(\varphi) $, above the whole space, respectively.
Thus,  $k(x,t)$, $u(x,t) $ , and $p(x,t)$ are called the kinetic, gradient, and potential energy densities, respectively.
The details of any collision can be more clarified by studying the evolution of all these parts throughout the collision process. Moreover, the sum of these terms (i.e. $\varepsilon=k+u+p$) is called the energy density function.
In particular, for the $\varphi^4$ system, the energy density function for the moving kink (antikink) solution (\ref{yh}), can be easily obtained as $\varepsilon=\gamma^2\sech^4(\gamma(x-x_{o}-vt))$.

In general, the velocity and rest mass ($m_{o}$) of any entity with a localized energy density can be obtained by the relativistic relations $v=P/E$ and $E^2=P^2+m_{o}^2$, respectively. Hence, for relativistic kink-bearing systems (\ref{fv}), if one can calculate the energy ($E$) and momentum ($P$) of each distinct entity (such as a kink) numerically, one can obtain its center of mass velocity as well as its rest mass using the above-mentioned relations.

\section{Numerical considerations}\label{sec3}

Since the dynamical equation (\ref{dy}) is non-linear, the superposition of two or multiple kinks
and antikinks is not necessarily a new solution. However, for $B\varphi^4$ systems, an alternative superposition of the kink and antikink solutions at initial times ($t\longrightarrow
0$) can be considered approximately a new solution provided that they are far enough from each other:
\begin{equation}\label{7}
	\varphi(x,t)=\sum_{i=1}^{m}(-1)^{i+1}\tanh[\gamma_{i}(x-v_{i}t-a_{i})]+C, \quad a_{i+1}-a_{i}\gg1,
\end{equation}
where $C=-1$ ($C=0$) if $m$ is an even (odd) number, $v_{i}$ is the initial velocity of the $i$th entity, and $a_{i}$ is the initial position of the $i$th entity (a kink or an antikink depending on whether $i$ is odd or even, respectively).
With constant $C$, this combination is constructed in such a way that the resulting field ($\varphi$) varies between two degenerate minima ( $-1$ and $+1$). In other words, the field must be equal to one of the vacua ( $-1$ and $+1$) at the boundaries ($x=\pm\infty$).
To ensure that the overlap of the kinks and antikinks is negligibly small, the initial relative distance ($a_{i+1}-a_{i}$) between alternative kinks and antikinks must be quite large. Accordingly, the energy density function corresponding to combination (\ref{7}) is seen as $m$ distant lumps, each moving with its own initial velocity ($v_{i}$):
\begin{equation}\label{8}
\varepsilon=\sum_{i=1}^{m}\gamma_{i}^2\sech^4(\gamma_{i}(x-a_{i}-v_{i}t)), \quad a_{i+1}-a_{i}\gg1.
\end{equation}
To put it differently, Eq.~(\ref{7}) can be expressly considered as a multi-particle-like solution.

In contrast to the well-known SG system, the other kink-bearing systems are not integrable. Thus, it is impossible to obtain their multi-solitonic solutions analytically. In other words, it is not possible to study the fate of any kink-antikink collision analytically.
Hence, to study collisions, it is necessary to use approximate methods (such as the collective coordinate approximation\cite{phi4cc,phi4cc2,phi62,CC2,CC3,CC4}  and Manton’s method \cite{Man1,Man2,Man3}) or quasi-exact numerical methods. This paper uses a numerical scheme with good accuracy based on the method of lines \cite{meofli} and the Runge-Kutta method. Thus, the above combination (\ref{7}) is a suitable initial condition for the numerical method.

\begin{figure}[ht!]
	\centering
	\includegraphics[width=150mm]{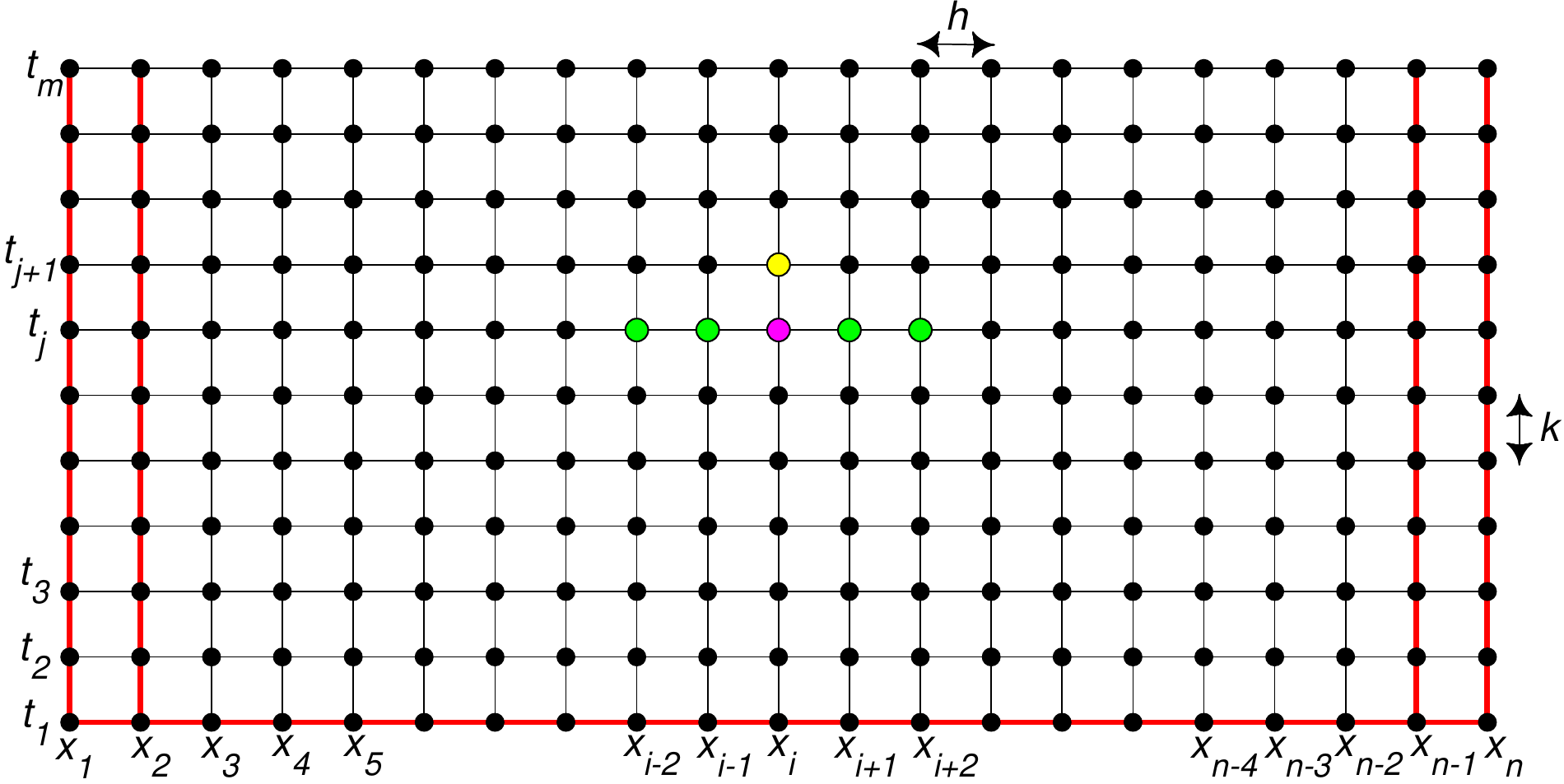}
	\caption{A grid of points on a space-time plane is considered. The temporal and spatial
		steps are indicated by $\textrm{k}$ and $\textrm{h}$, respectively. The second-order spatial derivative at any
		point (the purple point) can be approximately written as a linear function (\ref{vf}) of the field at the point itself and its four adjacent points (the green points). Using a Runge-Kutta
		method, the field value at the yellow point can be linked to the field value at the purple and green 	points. The field values for 	the points on the red lines are determined using initial and boundary conditions.
		} \label{mesh}
\end{figure}


The method of lines is based on the discretization of the spatial derivatives of some wave equations to obtain a set of coupled ordinary differential equations (ODEs) in terms of time. A standard method such as the Runge-Kutta method should be used to solve these coupled ODEs. The approximate discretized version of the second-order spatial derivative in the wave equation (\ref{dy}) can be written as:
\begin{equation} \label{vf}
		\left.\dfrac{\partial^2\varphi}{\partial x^2}\right|_{(x_{i},t_{j})}\approx\frac{1}{12\textrm{h}^2}(16\varphi_{i-1,j}-\varphi_{i-2,j}+16\varphi_{i+1,j}-\varphi_{i+2,j}-30\varphi_{i,j}),
\end{equation}
where $\varphi_{i,j}=\varphi(x_{i},t_{j})$ and $\textrm{h}$ is the spatial step size.
For such discretization  (\ref{vf}), the  magnitude of error  is  in the order of  $\textrm{h}^4$.  Therefore, the dynamical equation (\ref{dy}) can be approximately written   as:
 \begin{eqnarray}\label{ddy}
	&&\ddot{\varphi_{i}}=\dfrac{\partial^2\varphi_{i}}{\partial t^2}\approx H\left(\varphi_{i-2},\varphi_{i-1},\varphi_{i},\varphi_{i+1},\varphi_{i+2}\right)\nonumber\\&&\label{ddy2} \quad\quad\quad=\frac{1}{12\textrm{h}^2}(16\varphi_{i-1}-\varphi_{i-2}+16\varphi_{i+1}-\varphi_{i+2}-30\varphi_{i})-\frac{dV(\varphi_{i})}{d\varphi_{i}}.
\end{eqnarray} 	
For simplicity, we drop the index $j$. It should be noted that, according to Fig.~\ref{mesh}, we use
a two-dimensional space-time scheme containing $n\times m$ nodes for which the spatial and temporal step sizes are $\textrm{h}$ and $\textrm{k}$, respectively.
For any fixed spatial point ($x_{i}=x_{1}+(i-1)\textrm{h}$ ($i=1,\cdots,n$)), a one-parameter function $\varphi_{i}(t)=\varphi(x_{i},t)$ can be defined.
Hence, Eq.~(\ref{ddy}) can be considered as several coupled nonlinear second-order ODEs for time-dependent functions ($\varphi_{i}$).

According to Eq.~(\ref{vf}), it is clear that such symmetric discretization cannot be applied
to the points on the first, second, $(n-1)$th, and $n$th columns in the grid (the vertical red lines) presented in Fig.~\ref{mesh}.
Thus, the initial conditions and boundaries must be
adjusted to ensure that the values of the field at the boundaries remain approximately
constant ($\pm 1$) during the execution of the numerical program.
In other words, the boundaries must be far enough away from the central point of collision ($x=0$) to ensure that the effects of the collisions do not reach there during the running of the program.
For example, in this paper, to investigate a head-on kink-antikink collision (the kink and the antikink are initially at $a_{1}=-20$ and $a_{2}=20$, respectively, for which $v_{1}=-v_{2}=v$), the proper initial condition is considered as follows:
 \begin{equation}\label{tg}
 	\varphi(x,t)=\tanh[\gamma(x-vt+20)]-\tanh[\gamma(x+vt-20)]-1,\quad\quad t\longrightarrow 0,
 \end{equation}
To prevent the reflective properties of the boundaries from affecting the accuracy of the simulations, we set $x_{1}=-200$,  $x_{n}=200$, and $\varphi_{1}=\varphi_{2}=\varphi_{n-1}=\varphi_{n}=-1$ during the execution time ($t>400$).
Moreover, in this  paper, for the systems $0.2\leq B \leq 5$ ($B>5$), all the simulations were carried out by considering $\textrm{h} = \textrm{k} = 0.02$ ($\textrm{h} = \textrm{k} = 0.01$).




To use the Runge-Kutta method, it is first necessary to define the functions $\xi_{i}=\frac{d\varphi_{i}}{dt}$ ($i=3,\cdots n-2$). Therefore, the $n-4$ second-order coupled ODEs (\ref{ddy}) can be transformed into $2(n-4)$ first-order coupled ODEs:
\begin{equation}\label{kd}
	\begin{cases}
		\frac{d\varphi_{i}}{dt}=\xi_{i},~~\quad\quad\quad\quad\quad\quad\quad\quad\quad\quad\quad\quad i=3,\cdots, n-2,
		\\\frac{d\xi_{i}}{dt}=H\left(\varphi_{i-2},\varphi_{i-1},\varphi_{i},\varphi_{i+1},\varphi_{i+2}\right),\quad i=3,\cdots, n-2.	
	\end{cases}
\end{equation}
In general, if Eq.~(\ref{7}) at $t=0$ is considered as the initial condition for the field ($\varphi$), the initial condition for the function $\xi$ will be:
\begin{equation}\label{vin}
	\xi(x,t=0)=\sum_{i=1}^{N}v_{i}\gamma_{i}(-1)^{i}\sech^2[(-1)^{i+1}\gamma_{i}(x-a_{i})], \quad a_{i+1}-a_{i}\gg1.
\end{equation}
It should be noted that the function $\xi$ is almost zero at the points on the vertical red lines in Fig.~\ref{mesh}. However, using a fourth-order Runge-Kutta scheme with a small time-step $\textrm{k}$ for $ 2(n-4)$ first-order ODEs (\ref{kd}) leads to a quasi-exact numerical solution as follows:
\begin{eqnarray} \label{rk}
	&&\varphi_{i}(t_{j+1})=\varphi_{i}(t_{j})+\frac{1}{6}(K_{i1}+2K_{i2}+2K_{i3}+K_{i4}),\\&&\label{bf1}
	\xi_{i}(t_{j+1})=\xi_{i}(t_{j})+\frac{1}{6}(L_{i1}+2L_{i2}+2L_{i3}+L_{i4}),
\end{eqnarray}
in which
\begin{equation}\label{ax}
	\begin{cases}
		L_{i1}=\textrm{k}\xi_{i}(t_{j}),\quad
		\\K_{i1}=\textrm{k}H\left.\left(\varphi_{i-2},\varphi_{i-1},\varphi_{i},\varphi_{i+1},\varphi_{i+2}\right)\right|_{t=t_{j}},	
		\\	
		L_{i2}=\textrm{k}\bar{\xi}_{i},
		\\K_{i2}=\textrm{k}H\left(\bar{\varphi}_{i-2},\bar{\varphi}_{i-1},\bar{\varphi}_{i},\bar{\varphi}_{i+1},\bar{\varphi}_{i+2}\right),
			\\	
		L_{i3}=\textrm{k}\bar{\bar{\xi}}_{i},
		\\K_{i3}=\textrm{k}H\left(\bar{\bar{\varphi}}_{i-2},\bar{\bar{\varphi}}_{i-1},\bar{\bar{\varphi}}_{i},\bar{\bar{\varphi}}_{i+1},\bar{\bar{\varphi}}_{i+2}\right),
		\\
		L_{i4}=\textrm{k}\bar{\bar{\bar{\xi}}}_{i},
		\\K_{i4}=\textrm{k}H\left(\bar{\bar{\bar{\varphi}}}_{i-2},\bar{\bar{\bar{\varphi}}}_{i-1},\bar{\bar{\bar{\varphi}}}_{i},\bar{\bar{\bar{\varphi}}}_{i+1},\bar{\bar{\bar{\varphi}}}_{i+2}\right),
	\end{cases}
\end{equation}
where
\begin{eqnarray} \label{ed}
	&&\bar{\xi}_{i}=\xi_{i}(t_{j})+\frac{L_{i1}}{2},\quad \bar{\varphi}_{i}=\varphi_{i}(t_{j})+\frac{K_{i1}}{2},\\&&\label{ed2}
	 \bar{\bar{\xi}}_{i}=\xi_{i}(t_{j})+\frac{L_{i2}}{2},\quad \bar{\bar{\varphi}}_{i}=\varphi_{i}(t_{j})+\frac{K_{i2}}{2},\\&&\label{ed3}
	 \bar{\bar{\bar{\xi}}}_{i}=\xi_{i}(t_{j})+L_{i3},\quad \bar{\bar{\bar{\varphi}}}_{i}=\varphi_{i}(t_{j})+K_{i3}.
\end{eqnarray}
In the above equation, $i=3,\cdots,n-2$ and $j=1,\cdots,m-1$.

\section{Numerical Results}\label{sec5}

Parameter $B$ has practically no role in the details of a single kink (antikink). However, it is seen numerically  that when a kink interacts with an anti-kink, this parameter plays a very important role in the details of the interaction. In this section, our goal is to analyze numerically the effects of parameter $B$ on different quantities in the kink-antikink collisions. A proper initial condition for this goal could be the same as that introduced in Eq.~(\ref{tg}). The initial relative distance of $40$ between the kink and antikink is quite large; therefore,  the effects of their overlap are exponentially suppressed and negligible.

\subsection{The escape  windows}

The kink and antikink can have two possible fates. First, for some initial velocities, they
will bounce back and reflect from each other. Second, they will stick together and a bion state will be generated. A bion is a long-lasting non-topological oscillating bound state that decays slowly via emitting its energy in the form of small-amplitude waves. For some special systems (the so-called radiative systems), the kink-antikink collision will lead to immediate annihilation \cite{RD,RD1}.

 \begin{figure}[ht!]
	\centering
     \includegraphics[width=140mm]{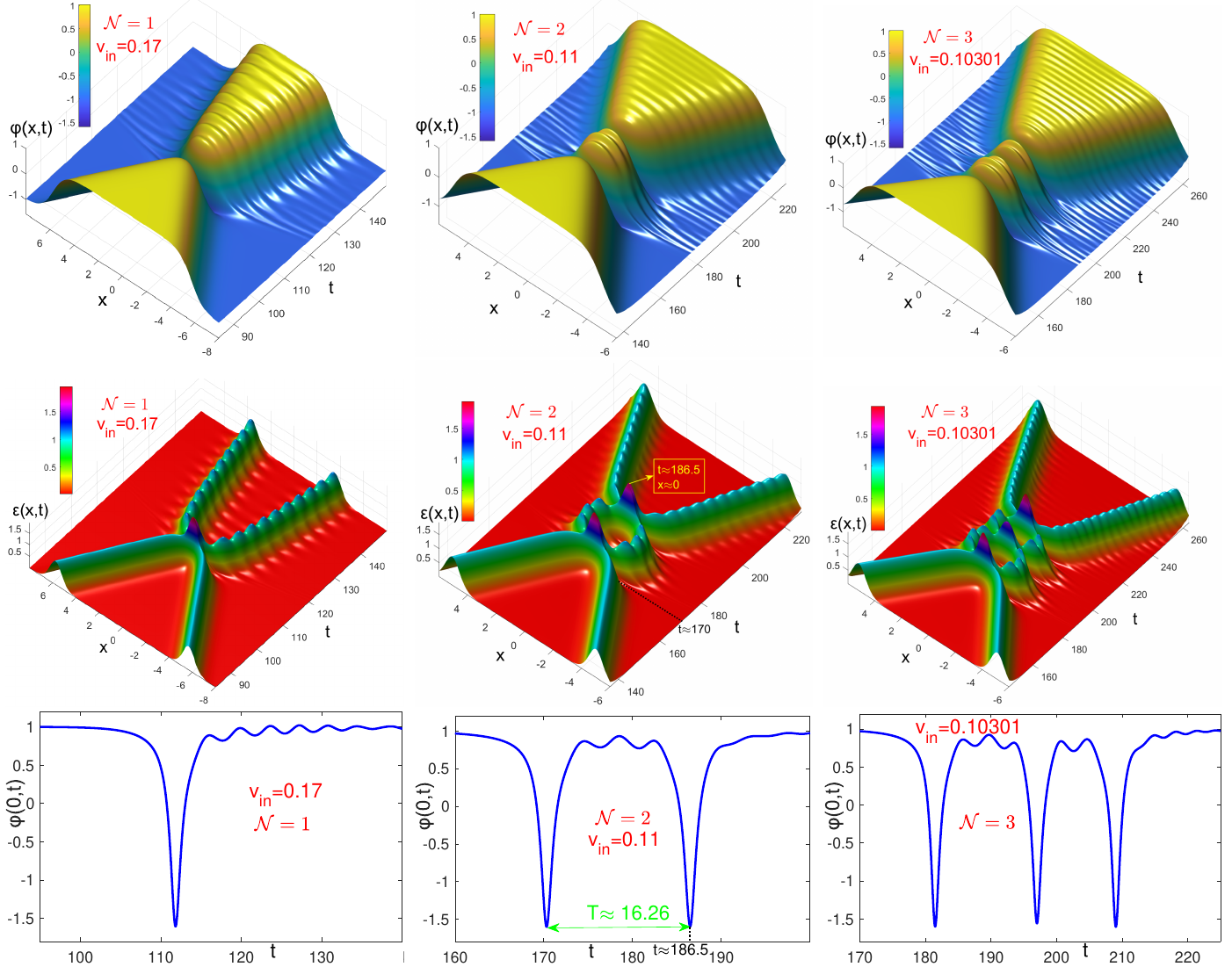}
	\caption{For the system $B=1.4$, the first (second) row shows the field (energy density) evolution for a one-bounce, two-bounce, and three-bounce collision, respectively, at different initial velocities. The third row shows the field variations in the center of mass for the collisions drawn in the first row.} \label{mnj}
\end{figure}

In all studies that have been conducted on kink-bearing systems (except for SG and radiative systems), a specific velocity (or the so-called critical velocity ($v_{cr}$)) has always been found whether they have non-trivial internal modes or not. It is well-known that for the initial velocities greater than $v_{cr}$, the kink and antikink do not stick to each other in a head-on collision. For the initial velocities below the critical velocity, the predominant phenomenon is the sticking together of the kink and anti-kink and the generation of a bion state.
However, there are numerous wide and narrow intervals of the initial velocity below the critical velocity at which the kink and antikink finally escape after a finite number of collisions.
   These special intervals are called escape windows, and the entire range of initial velocities that includes all of them can be called  the escape-window region.
 Depending on how many times the kink and antikink collide before escaping, 
such escape windows are classified into two-bounce, three-bounce, four-bounce and so on.
Consequently, an ${\cal N}$-bounce escape  window is an interval of initial velocities at which the kink and antikink collide ${\cal N}$ times before escaping.
For example, let us consider the case of $v_{in}=0.11$ in Fig.~\ref{mnj}, which is a member of a two-bounce escape window. As it is seen,  kink-antikink collides at $t\approx170$,  bounces back, recedes to finite separation, and then collides again at $t\approx186.5$.  
This phenomenon is described as the resonance energy exchange between the translational and vibrational internal modes and will occur in any ${\cal N}$-bounce collision (resonance) \cite{DSG1,DSG2,phi44,phi47,phi46,phi61}.

 \begin{figure}[ht!]
	\centering
	\includegraphics[width=120mm]{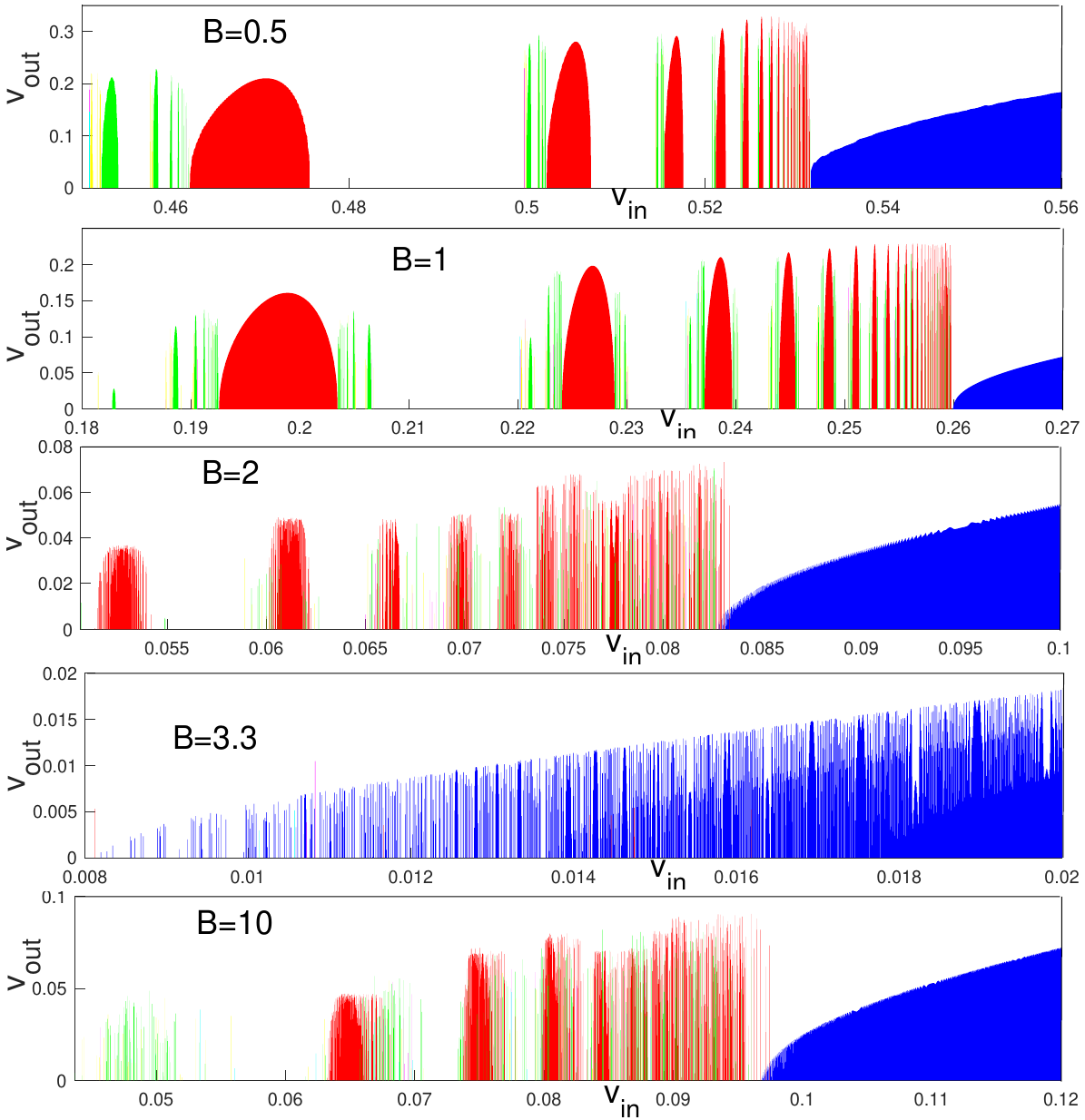}
	\caption{Bar representation showing the output velocity versus the initial velocity for five different   systems. Interesting details can be observed by zooming in on the bars (especially the blue ones). Except for the case $B = 1$ ($\varphi^4$ systems), the ends of the bars do not form a smooth curve} \label{voutvin}
\end{figure}

\begin{figure}[ht!]
	\centering
	\includegraphics[width=132mm]{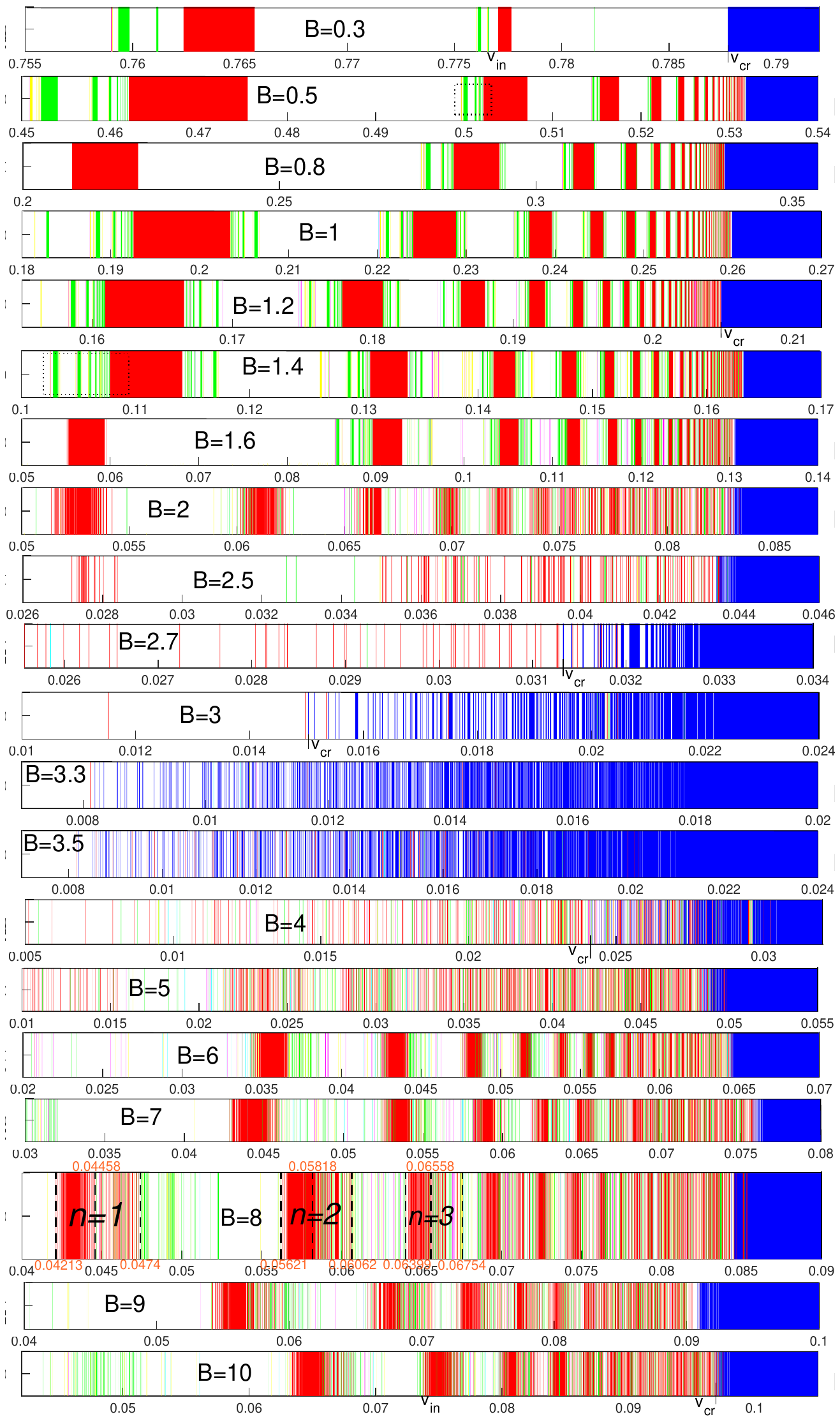}
	\caption{The variety of different fates versus the incoming (initial) velocity for various systems. The white bars show the initial velocities that lead to bion states. The blue, red, green, yellow, purple, and cyan bars represent the collisions of one-bounce, two-bounce, ..., and six-bounce, respectively. The $v_{in}$-axes are divided into steps of $0.00001$.} \label{norm}
\end{figure}

To illustrate the details of the escape  windows in different systems,
a high-resolution colored diagram of the final velocity in terms of the initial velocity with
a small step size of $0.00001$ was constructed.
Some of these diagrams are shown in Fig.~\ref{voutvin} in which the blue, red, green, yellow, purple, and cyan bars represent the one-bounce, two-bounce, ..., and six-bounce collisions, respectively. The height of each bar indicates the final velocity. In regions with no colored bars (white bars), the collisions lead to the formation of a bion state. The non-zero bars can be normalized to one to better show the various escape  windows for many systems (see Fig.~\ref{norm}).
In this regard, the two-bounce escape  windows are the intervals of the initial velocities corresponding to the continuous joining of the red bars. Typically, there should not be even one thin bar with a different color among the bars of a special escape  window.  However, comparing different $v_{out}-v_{in}$ diagrams shows that this expectation is not a general statement.
For example, according to Fig.~\ref{norm}, many intervals of the red bars (two-bounce escape  windows) are intersected by some thin bars of different colors for the cases in which $B>1$.
Furthermore, for the systems close to $B=3.3$, the
distinct escape  windows completely disappear and a chaotic situation emerges. In other words, the escape  windows become fuzzier as $B$ gets closer to $3.3$. Hence, the constraint on the continuous joining of the bars with the same color for a special escape  window is violated when $B>1$.

In general, all  ${\cal N}$-bounce escape windows can be classified based on the similarity in collision. By ‘similarity in collision’, we mean having similar details for field variations in collisions,  which will be examined in detail in subsection \ref{Time cycles}  for  two-bounce escape windows.  For example,  the first three classes of two-bounce  escape windows (red bars) are indicated in Fig.~\ref{norm} for the case of $B=8$. 
It should be noted that  the members of a class may contain several separate initial velocity intervals.
In other words, we no longer the continuous joining of the bars with the same color  as a criterion for the classification of ${\cal N}$-bounce escape windows, although that is still  valid for $B\leq 1$.

Overall, the presence of an escape-window region, which can  usually be  expected to be seen before the critical speed, seems to be a common routine  in all systems. Depending on  parameter $B$, this region,  sometimes occurs for very low initial velocities (even smaller than in the $\varphi^6$ case \cite{phi61}) and sometimes starts at  very high velocities. Of course, for systems close to $B=0.2$, the complexity of this region (such as the case of $B=0.3$ in Fig.~{\ref{norm}) is decreased, and for those that are very close to $B=0.2$, this region disappears. 
Accordingly, if the resonance energy exchange phenomenon can be considered as the only phenomenon involved in the formation of an escape-window region,  the resonance frequency\footnote{It will be introduced in subsection \ref{Time cycles}.}  is expected to significantly depend on $B$.  Contrary to this expectation, it will be shown that the resonance frequency is almost independent of $B$.
Hence, it seems that there are more factors that are effective in the occurrence of  such escape-window regions. In this regard, a more detailed investigation can be of interest to researchers in future studies.	
For example, a modified version of the collective coordinate method (CCM)  based on Derrick modes, which have been recently used to describe the kink-antikink collisions in the SG and $\varphi^4$ model \cite{phi4cc2}, may be useful to explain such escape-window regions obtained in this paper.




\subsection{The critical velocity}

One of the significant features of the critical velocity is that one-bounce collisions do
not occur at initial velocities less than it. In addition, only one-bounce collisions are expected to be seen for initial velocities larger than the critical velocity. In other words, it is expected that no bars with other colors are among the blue bars and vice versa and that the first special initial velocity after which the blue bars appear is the critical velocity.
Accordingly, preparing a diagram of the output velocity versus the initial velocity (and obtaining different
${\cal N}$-bounce windows) is the standard way to obtain the critical velocity (Figs.~\ref{voutvin} or \ref{norm}). In this regard, the cases $B =0.3$ and  $B =1.2$  in Fig.~\ref{norm} can be mentioned.
This is the way to find the critical velocity. However, examining such diagrams for the entire range of $B$ (from $1$ to $10$) shows that these expectations are not always true. In other words, the blue and other colored bars are intermingled in the systems in which $B$ is approximately larger than $1.6$. Therefore, it is practically impossible to determine the critical velocity based on what we have accepted so far.

\begin{figure}[ht!]
	\centering
	\includegraphics[width=150mm]{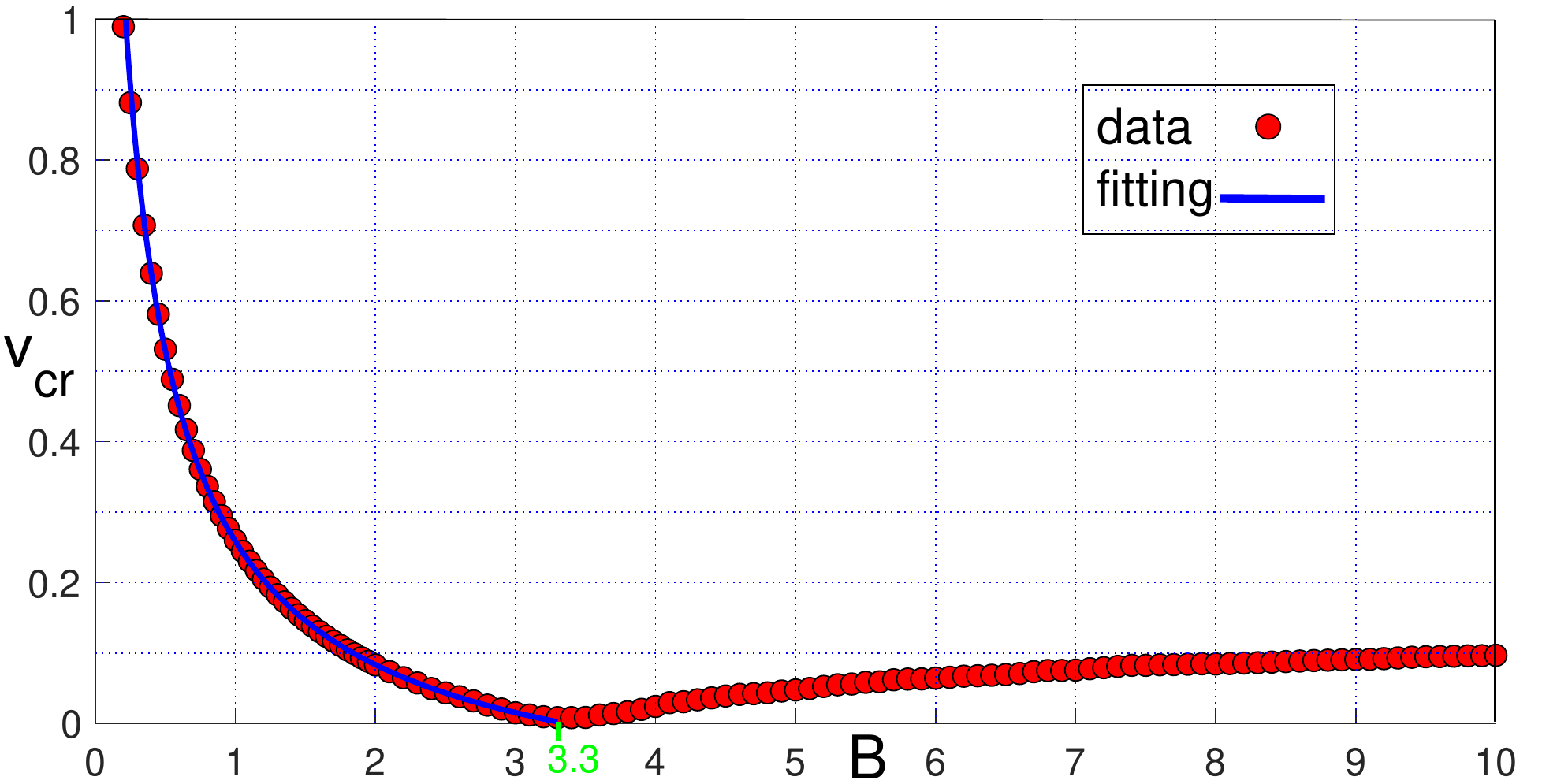}
	\caption{The critical velocity versus $B$ obtained from the numerical calculations (solid red circles) and fitted by Eq.~\ref{ans} in the range of $0.2\leq B \leq 3.3$ (the blue curve). The minimum occurs at about $B=3.3$. To obtain a smooth curve for $B>2$, some points are slightly shifted, i.e. $v_{cr}\longrightarrow v_{cr}+\delta v_{cr}$, where $\delta v_{cr}\leq 0.001$.} \label{vcrvsB}
\end{figure}




Thus, to obtain a $v_{cr}-B$ diagram, we have to give a new definition of the critical
velocity. We can define the critical velocity as a specific initial velocity at which the first
one-bounce collision occurs. For example, based on this new definition, the critical velocity
for the case $B=4$ is about $v_{cr}=0.02414$ according to Fig.~\ref{norm}.
Accordingly, the critical velocity is numerically obtained as a function of $B$, leading to Fig.~\ref{vcrvsB}. It was observed that $v_{cr}$ increases with the decrease of $B$ so that it tends to the velocity of light ($v_{cr}\longrightarrow 1$) for the values of $B$ that are less than $0.2$.
Furthermore, according to Fig.~\ref{vcrvsB}, the numerical results show that the minimum critical velocity occurs at about $B=3.3$ which is equal to $v_{cr}\approx0.0082$. For $B>3.3$, the critical velocity exhibits a different behavior as a function of $B$ in such a way that it grows monotonically.
For the interval $0.2\leq B \leq 3.3$, we guess the following ansatz:
\begin{equation}\label{ans}
	v_{cr}(B)\approx\ln (2+B^{-0.95})-0.84,
\end{equation}
which is fitted with the obtained numerical results with a good confidence level (see the blue curve in Fig.~\ref{vcrvsB}). Unfortunately, we did not succeed in finding another proper formula to fit the numerical results for the range of $B>3.3$.

If we assume two distinct forces of attraction and repulsion in the collision between kink and anti-kink, according to Fig.~\ref{vcrvsB}, we can conclude that the repulsion force tends to zero for small values of $B$, because $v_{cr}$ tends to $1$, and a bion state is always created.   In other words, the more horizontal  the shape of the potential  in region $|\varphi|>1$, the less strong this force. Hence, on the contrary, it can be expected that this force will increase with the increase in $B$ or in the verticality of   the potential curve in $|\varphi|>1$. This impression is true for $B<3.3$, but for $B>3.3$, it is contrary to expectations. In the current situation, we do not have an explanation for the occurrence of this unexpected behavior, but the presentation of a model based on the CCM \cite{phi47,phi4cc,phi4cc2} may shed more light on it. 







\subsection{Two-bounce escape  windows and time cycles} \label{Time cycles}

For any kink-bearing system (except for the SG and radiative systems), there are many two-bounce windows in the $v_{out}-v_{in}$ diagram whose width decreases as the initial velocity
increases. Incremental integers can be used to label (classify) two-bounce windows (i.e. the red intervals in Fig.~\ref{norm}) from left to right and as their orders.
The number of cycle oscillations (small oscillations of the field \cite{phi46}) between the first and second collisions for each two-bounce escape  window is an invariant characteristic (see Fig.~\ref{B8tc} and the case $B=8$ in Fig.~\ref{norm}). The cycle oscillations are introduced in terms of the field variations at the center of mass for any collision (i.e. $\varphi(x,0)$) as shown in Figs.~\ref{B8tc} and \ref{tcy}. For example, for the case $B=8$, all the initial velocities of the first, second, and third escape  windows respectively lead to four, five, and six cycle oscillations between two successive collisions.
It should be noted that the number of cycle oscillations in the first two-bounce window depends on parameter $B$ (see Figs.~\ref{tcy} and  \ref{NC}). As shown in Figs.~\ref{B8tc} and \ref{NC}, the ($n$)th two-bounce window has a longer time interval and one more cycle oscillation between the first and second collisions compared to ($n-1$)th two-bounce window.
Therefore, for a special system, the important feature for determining a special two-bounce escape  window is having the same cycle oscillations between successive collisions. Based on this, if a set of several separate initial velocity intervals leads to the same number of cycle oscillations, all together they form a special escape  window. 
Accordingly, the similarity in the details between successive collisions can be considered a basis for classifying ${\cal N}$-bounce escape windows regardless of the continuous joining of different initial velocity intervals.



\begin{figure}[ht!]
	\centering
	\includegraphics[width=140mm]{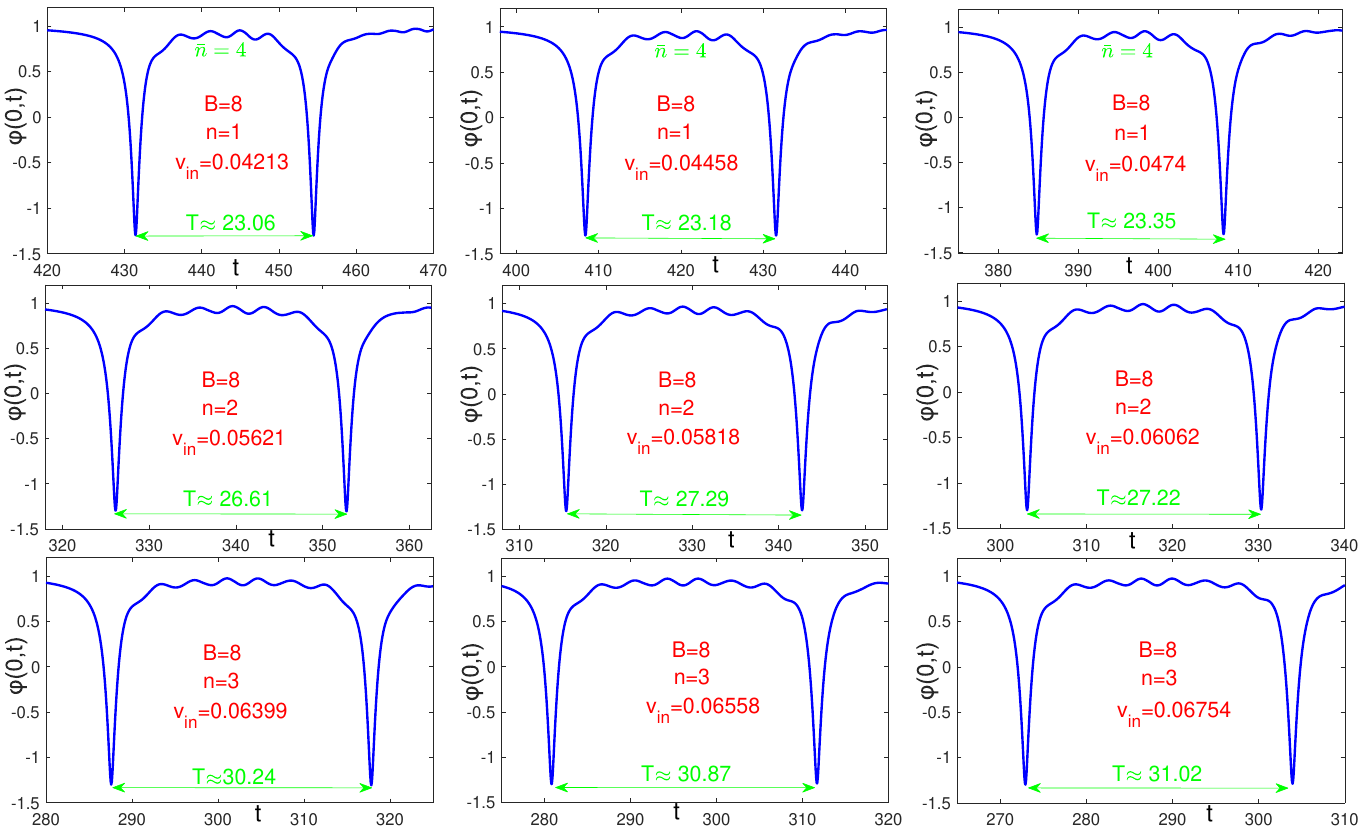}
	\caption{The field variations at the center of mass for different initial velocities belonging
		to the first, second, and third two-bounce escape  windows of the system $B=8$ (see
		Fig.~\ref{norm}).} \label{B8tc}
\end{figure}

\begin{figure}[ht!]
	\centering
	\includegraphics[width=140mm]{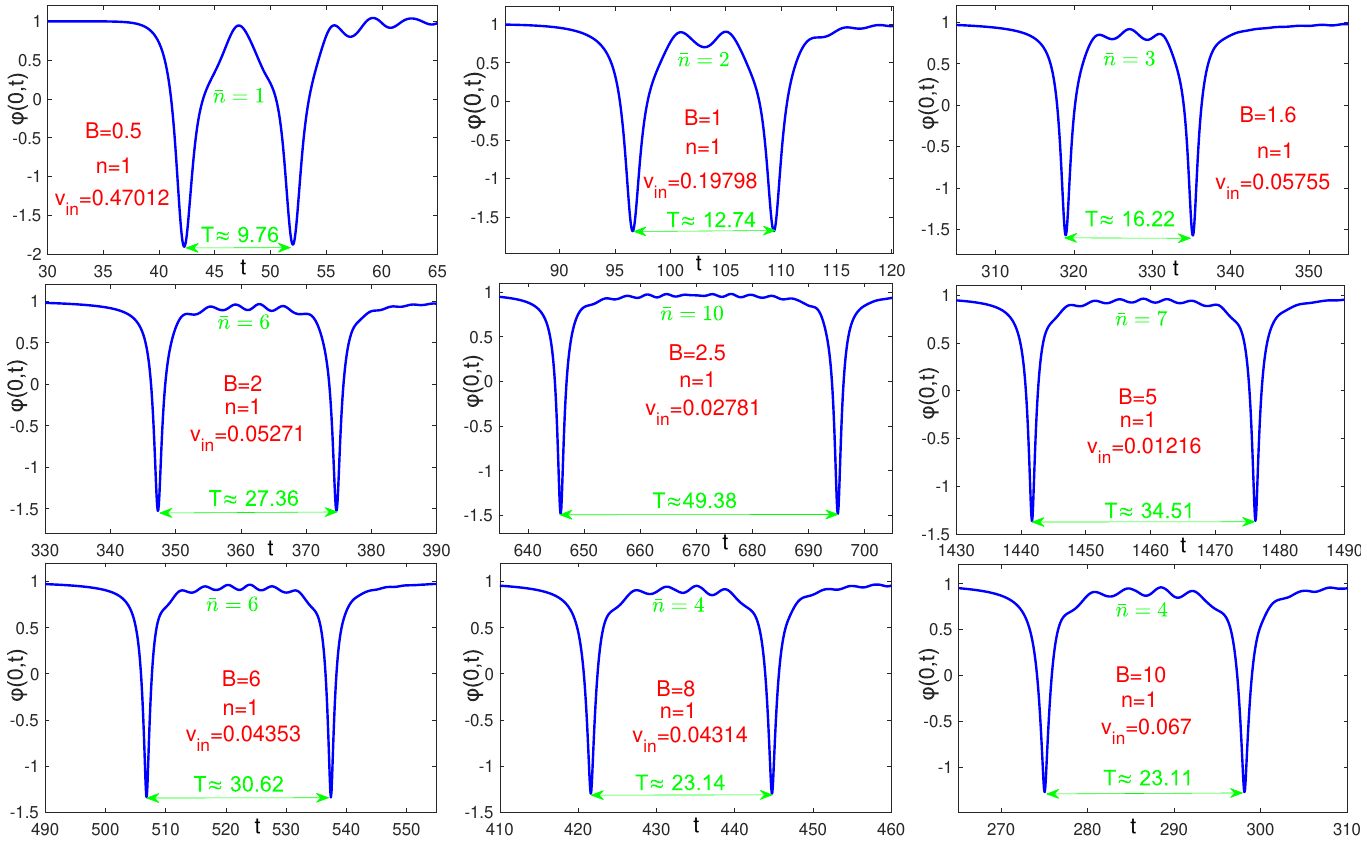}
	\caption{The field variations at the center of mass corresponding to the first two-bounce escape  windows of the different systems.} \label{tcy}
\end{figure}

\begin{figure}[ht!]
	\centering
		\includegraphics[width=140mm]{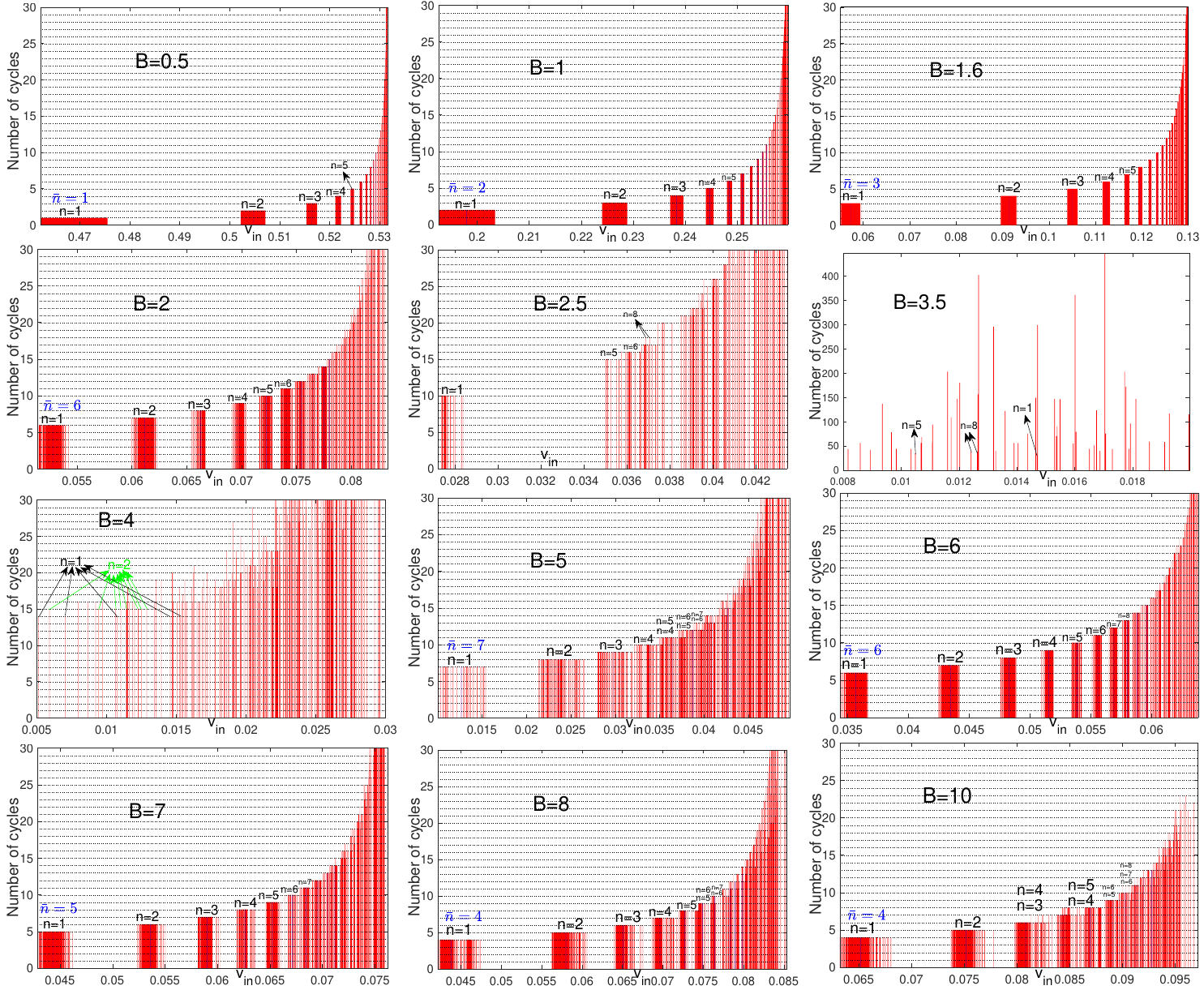}
	\caption{The number of cycles between two successive collisions in two-bounce escape 
		windows for different systems. A set of bars with the same height corresponds to a two-bounce escape  window with a special order. A set of bars with the lowest height is numbered as $n=1$. The other two-bounce windows are numbered based on their height difference from the first two-bounce window. The midpoints of some two-bounce escape  windows are indicated by blue bars.} \label{NC}
\end{figure}

Figure~\ref{NC} contains interesting and noteworthy points. For example, an incremental sequence of numbers is not observed for systems  close to $B=3.3$ as for the one with $B=2.5$ so that the second, third, and fourth two-bounce escape  windows are absent though they were expected to appear regularly between the first and fifth two-bounce escape  windows.
It is also clear that two-bounce escape  windows with different orders are not always separated as distinct islands. This is significantly intensified for $B$ values close to $3.3$ which are mixed together. Even the first two-bounce escape  windows are mixed with other ones. In addition, the height of the bars increases as $B$ gets closer to $3.3$.

\begin{figure}[ht!]
	\centering
	\includegraphics[width=120mm]{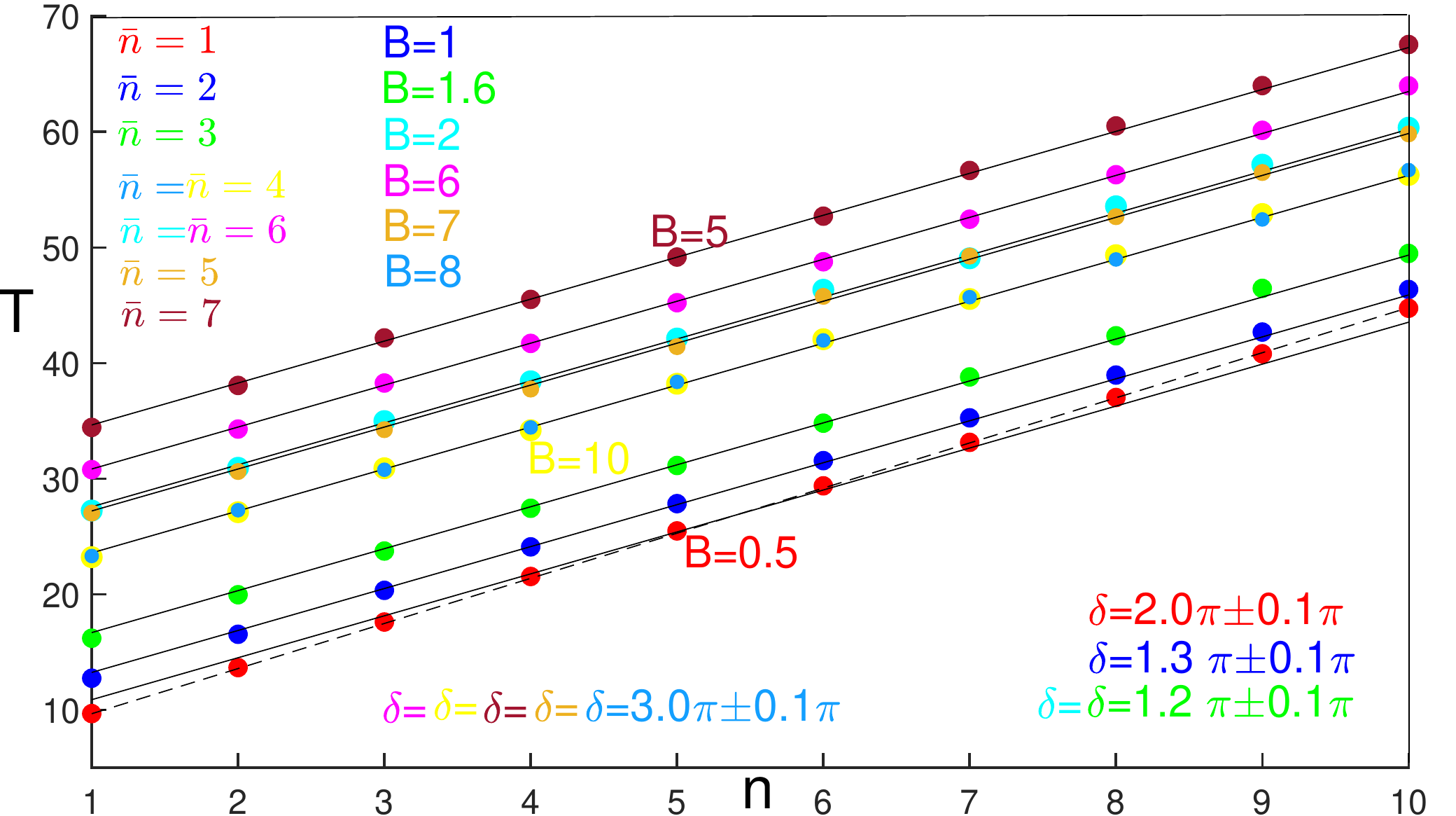}
	\caption{The time ($T$) between two collisions versus $n$ for the first ten two-bounce escape  windows. The solid circles indicate the results of our numerical calculations. The solid lines are the corresponding fitting of Eq.~(\ref{bn}) with the proper values of the phase ($\delta$). } \label{TC}
\end{figure}

\begin{figure}[ht!]
	\centering
	\includegraphics[width=120mm]{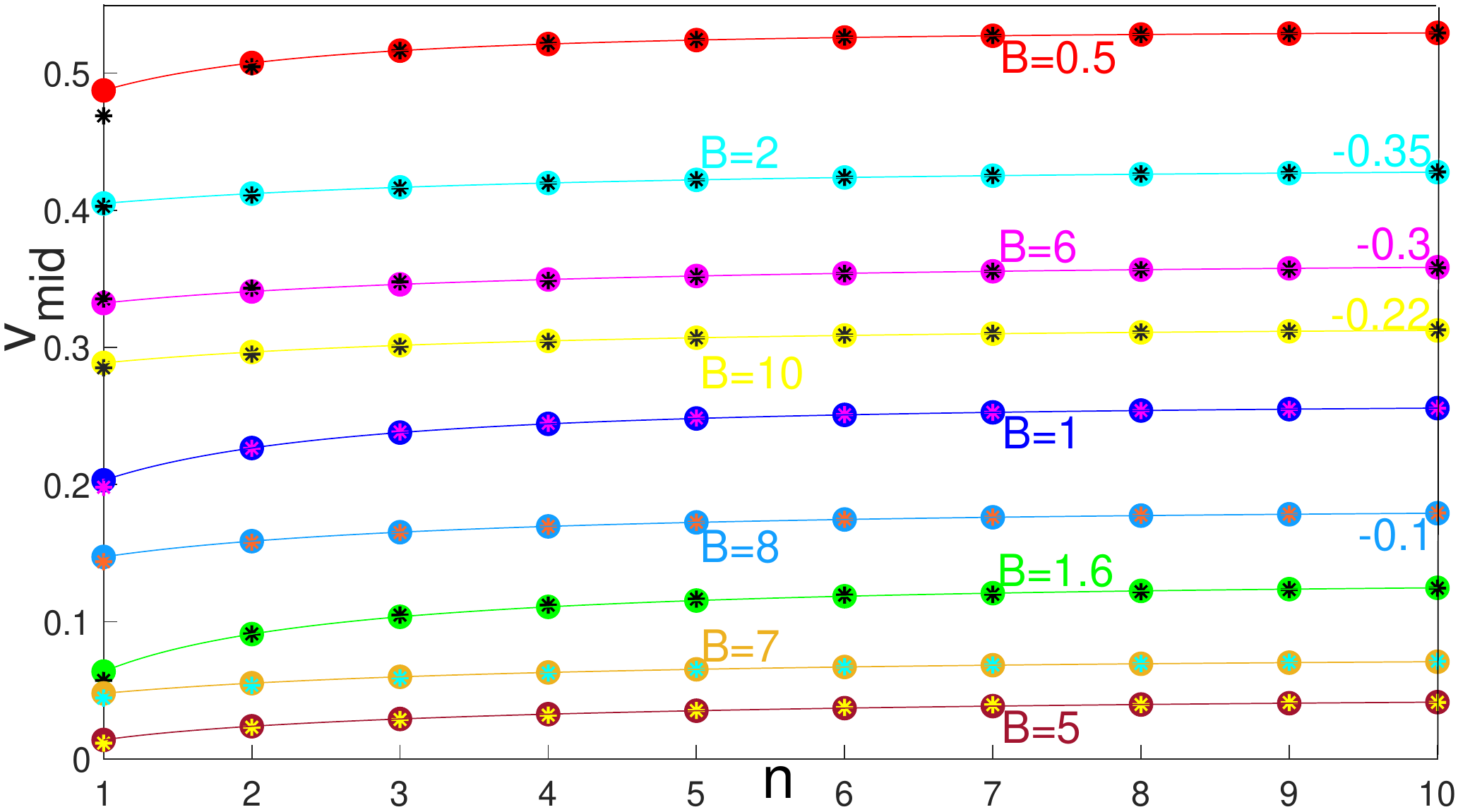}
	\caption{The initial velocity at the midpoint of the two-bounce escape  windows ($v_{mid}$)
		as a function of $n$ for different systems. The colored curves and solid circles are the estimations of equation (\ref{bn2}) and the stars show the results of our numerical calculations. To prevent the curves from overlapping and to make them clearly distinct, we had to shift some of them by certain values. To obtain the correct curves, the numbers written on each curve must be added. } \label{curves}
\end{figure}

\begin{figure}[ht!]
	\centering
	\includegraphics[width=140mm]{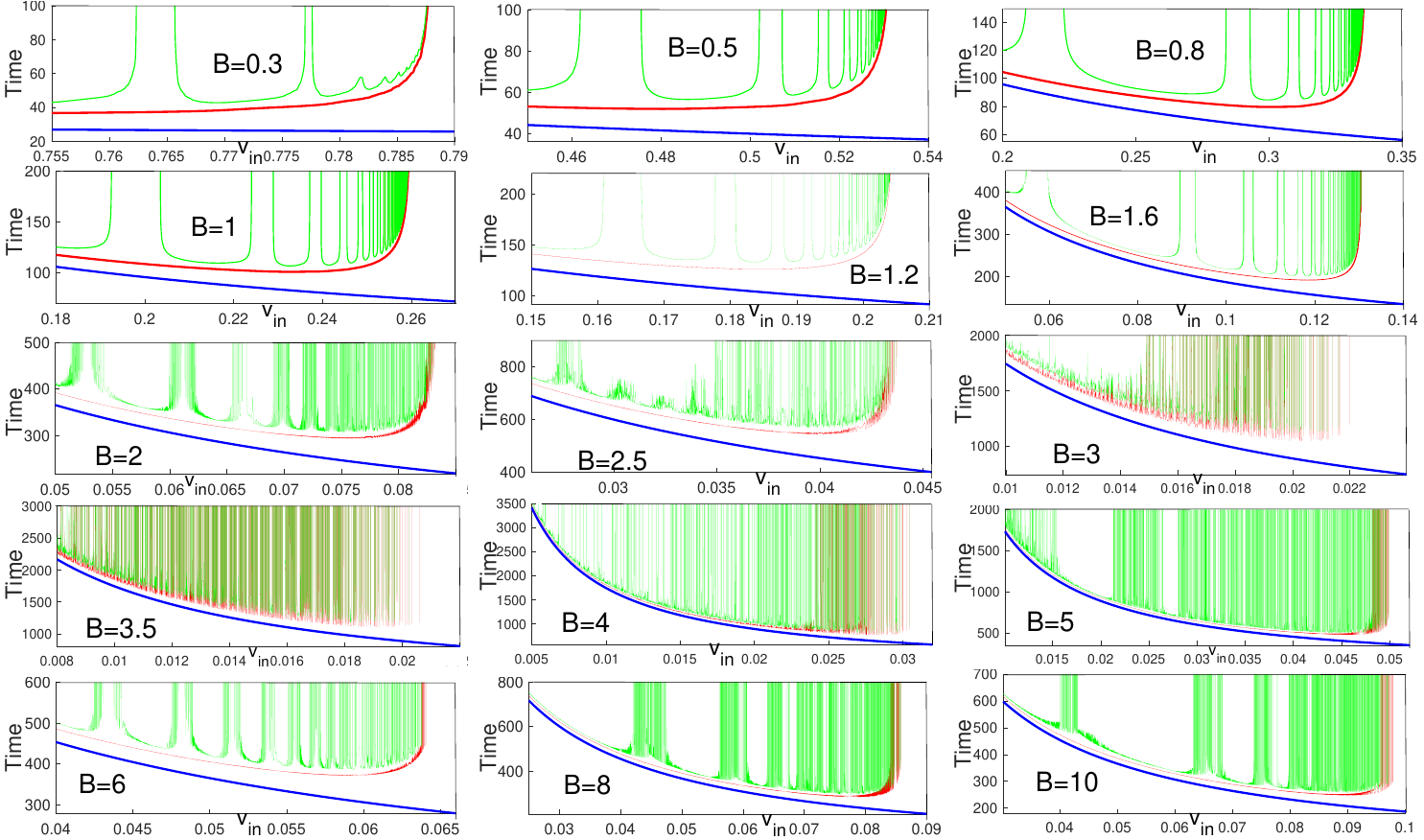}
	\caption{ The times of the first (blue), second (red), and third (green) kink-antikink collisions as a function of initial velocity for various systems. } \label{times}
\end{figure}

The elapsed time ($T$) between  two collisions in a two-bounce window   is not constant and increases slowly versus incoming velocity ($v_{in}$).
However, it is common to consider the midpoint of different two-bounce escape  windows to define some specific times ($T$) as a function of $n$ (see the blue bars in Fig.~\ref{NC}). There is a clear phenomenological relation that very well explains the relation of $T$ in terms of $n$ \cite{phi44,phi477,DSG1,phi81,book,polonomial3}:
\begin{equation} \label{bn}
\omega_{s}T=2\pi (n+\bar{n})+\delta
\end{equation}
where $\omega_{s}\approx\sqrt{3}$ is the resonance frequency, $n$ is the number of the consecutive two-bounce escape  windows in the sequence shown in Fig.~\ref{NC}, $\delta$ is a constant phase shift, and $\bar{n}$ is the number of cycle oscillations corresponding to the first two-bounce window ($n=1$)
which is presented in Fig.~\ref{tcy}. During a two-bounce collision, the kinetic energy of the kink
and antikink is transferred to the oscillating mode with frequency $\omega_{s}$ and then returns back
to the kinetic energy mode provided that there is a certain resonance relation (\ref{bn}) between $T$ and $\omega_{s}$.
In general, the resonance frequency ($\omega_{s}$) and the kink vibrational mode frequency ($\omega_{o}$) are not supposed to be precisely the same \cite{phi44,DSG1,DSG2,phi477,phi81}. However, they almost match in the $B\varphi^4$ systems introduced in this paper, i.e. $\omega_{s}\approx\omega_{o}$. Thus, by selecting some proper phases in equation (\ref{bn}), the points in Fig.~\ref{TC} are nicely fitted with some parallel lines.  
According to Fig.~\ref{TC}, it can be concluded that for $B\geq 5$ ($B\leq 2$), the proper phase is $\delta \approx 3\pi$ ($\delta\leq 2\pi$). Furthermore, the greatest deviation is seen for the system $B=0.5$. Thus, by fitting the best line (i.e. the dotted line), the resonance frequency value is obtained as $\omega_{s}\approx\sqrt{3}+0.12$.


As one result of the resonant energy exchange theory, another phenomenological relation
predicts the initial velocity at the midpoint of consecutive two-bounce escape  windows \cite{phi44,DSG1,phi81}:
\begin{equation} \label{bn2}
	v_{mid}^{th}(n)=\sqrt{v_{cr}^2-\dfrac{\omega_{s}^2\bar{\beta}^2}{ (2\pi (n+\bar{n})+\delta)^2}}
\end{equation}
 in which $\bar{\beta}$ is the mean value of
\begin{equation} \label{dfv}
\beta(n)=T(n)\sqrt{v_{cr}^2-v_{mid}^2(n)},
\end{equation}
where $v_{mid}$ is obtained numerically for the midpoint of each two-bounce window as shown
by blue bars in Fig.~\ref{NC}. Accordingly, Fig.~\ref{curves} shows the good correspondence between the $v_{mid}$-theoretical (solid circles) and $v_{mid}$-numerical (stars) predictions.
This correspondence is not high for $n = 1$. However, as $n$ increases, the proximity of the numerical and theoretical results is usually enhanced. Among the selected investigated systems, the greatest deviation between the numerical and theoretical results occurs for the case $B=0.5$. The validity of equations (\ref{bn}) and (\ref{bn2}) is significantly lost for systems close to $B=3.3$ due to their prevailing chaotic behavior.  
By comparing the obtained numerical results with the results obtained from the phenomenological relations (\ref{bn}) and (\ref{bn2}), it can be concluded  that the resonance frequency ($\omega_{s}$) is almost the same as the kink vibrational mode frequency ($\omega_{o}$), regardless  of parameter $B$.

Plotting the first, second, and third collision times versus the initial velocity is another way to depict two-bounce escape  windows. In this regard, the blue, red, and green curves in Fig.~\ref{times} respectively indicate the times until the first, second, and third collisions as a function of $v_{in}$.
The initial velocities at which the times of the second collision are finite and the those of the third collision become infinite belong to the two-bounce windows. An infinite time for the second collision indicates a one-bounce collision. Based on Figs.~\ref{norm}, \ref{NC}, and \ref{times}, an increase in chaotic behavior from $B=1$ to $B=3.3$ can be clearly seen.
For systems close to $B=3.3$, this chaotic behavior peaks. This behavior decreases when $B$ becomes larger than $3.3$ and a rather regular behavior is observed around $B=6$. When the value of this parameter becomes larger than $6$, a slowly increasing chaotic behavior is observed.


\subsection{The quasi-fractal structure}


\begin{figure}
	
	\includegraphics[width=150mm]{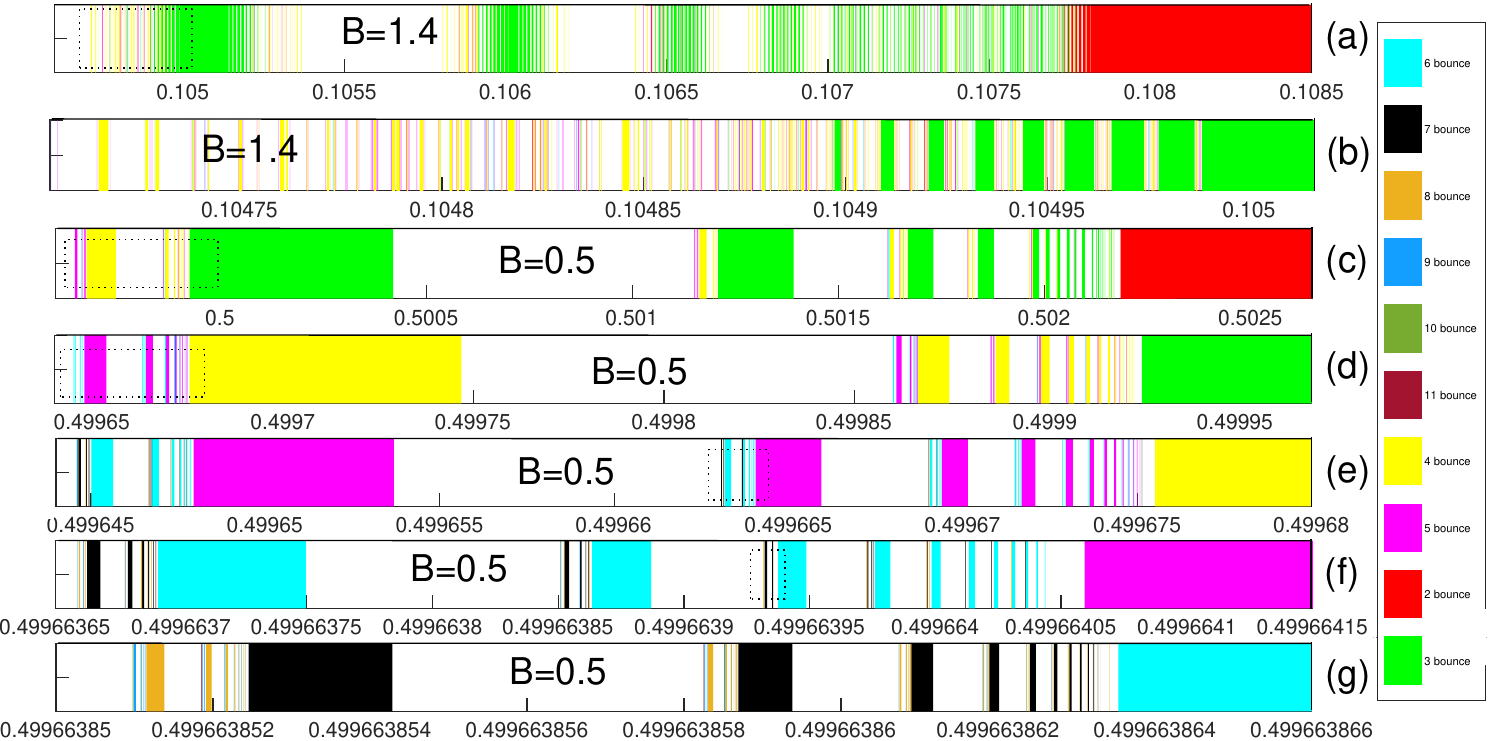}

	\caption{Each dotted rectangle shows a selected part of a subfigure that is plotted in
		more detail in the next subfigure with a smaller step size. The fractal structure is seen
		for the system $B=0.5$ but not for the system $B=1.4$. The first parts of the normalized $v_{out}-v_{in}$ diagrams which belong to the systems $B=0.5$ and $B=1.4$ are illustrated in Fig.~\ref{norm}.} \label{frac}
\end{figure}




The existence of a quasi-fractal structure is a well-known feature of escape  windows. In other words, each ${\cal N}$-bounce escape  window is surrounded by other (${\cal N}+i$)-bounce
escape  windows ($i=1,2,\cdots$) with smaller widths.
For example, if we consider the case
$B=0.5$ in Fig.~\ref{norm}, each red-colored (two-bounce) escape  window is surrounded (only from the left) by green, yellow, purple, and cyan escape  windows. Likewise, each
green-colored (three-bounce) escape  window is surrounded by yellow, purple, and cyan escape  windows.
In this regard, the distribution of lower-order escape  windows around each escape  window has a structure similar to the prototype (Fig.~\ref{norm} and \ref{frac}).  For example, if the outlined black dotted rectangular region in ~\ref{norm} (case $B=0.5$) is examined with more nodes, Fig.~\ref{frac}-(c) is obtained which sheds more light on this region.
Fig.~\ref{frac}-(c)  has a structure similar to that of Fig.~\ref{norm} for the case $B=0.5$ provided that the colors change in order (i.e. red to blue, green to red, yellow to green, and so on). In the next step, another outlined rectangular region can be considered in Fig.~\ref{frac}-(c) and zooming in on it with more nodes results in Fig.~\ref{frac}-(d).
Again, the result is similar to Fig.~\ref{norm} for the case $B=0.5$ provided that we change green to blue, yellow to red, blue to green, and so on. This routine is repeated for the yellow, purple and cyan intervals in the subsequent figures. What we can see here is the existence of a quasi-fractal structure which may be considered as a general rule for any ${\cal N}$-bounce escape  window.  In other words, the more we
repeat the zooming-in process, the narrower the intervals that will emerge with higher orders of escape  windows and similar structures.
Note that this rectangular region could be assumed on the right edge of any special escape  window. This is not the case for some systems such as $B=0.5$ which do not have any structure on the right edges of the escape  windows. However, in any case, the fractal structure is not expected to change.

This important feature of the escape  windows does not always exist in all the systems introduced in this article. In fact, for the systems $B>1$, this fractal structure obeys the rules only for a finite number of zooming in. As was mentioned above, the escape  windows gradually become fuzzy, the clear boundaries between them are gradually disturbed, and different escape  windows are mixed clearly for the systems close to $B=3.3$. Significantly, an obvious chaotic situation is observed for systems close to B=3.3 which means that the expectation of a quasi-fractal structure is completely out of the question.
For example, for the system $B=1.4$, although a structure similar to the prototype is obtained in the first zooming in on the edge of a two-bounce escape  window, the original structure is not repeated for the next zooms because of becoming  fuzzy (see the case $B=1.4$ in Fig.~\ref{norm} as well as  parts (a) and (b) in  Fig.~\ref{frac}).

Furthermore, the numerical results reveal that the higher-order ${\cal N}+i$ escape  windows are not necessarily symmetrically distributed around any ${\cal N}$-bounce escape  window. For instance, in the case $B=0.5$, they are all distributed to the left as shown in Fig.~\ref{norm}.  In the case $B=6$, they are mainly observed on the right. Moreover, an interesting and exceptional situation is observed for the case $B=0.8$ in which no higher-order windows are seen around the first two-bounce escape  window.



\subsection{The output  velocity as a function of  $B$}

\begin{figure}[ht!]
	\centering
	\includegraphics[width=130mm]{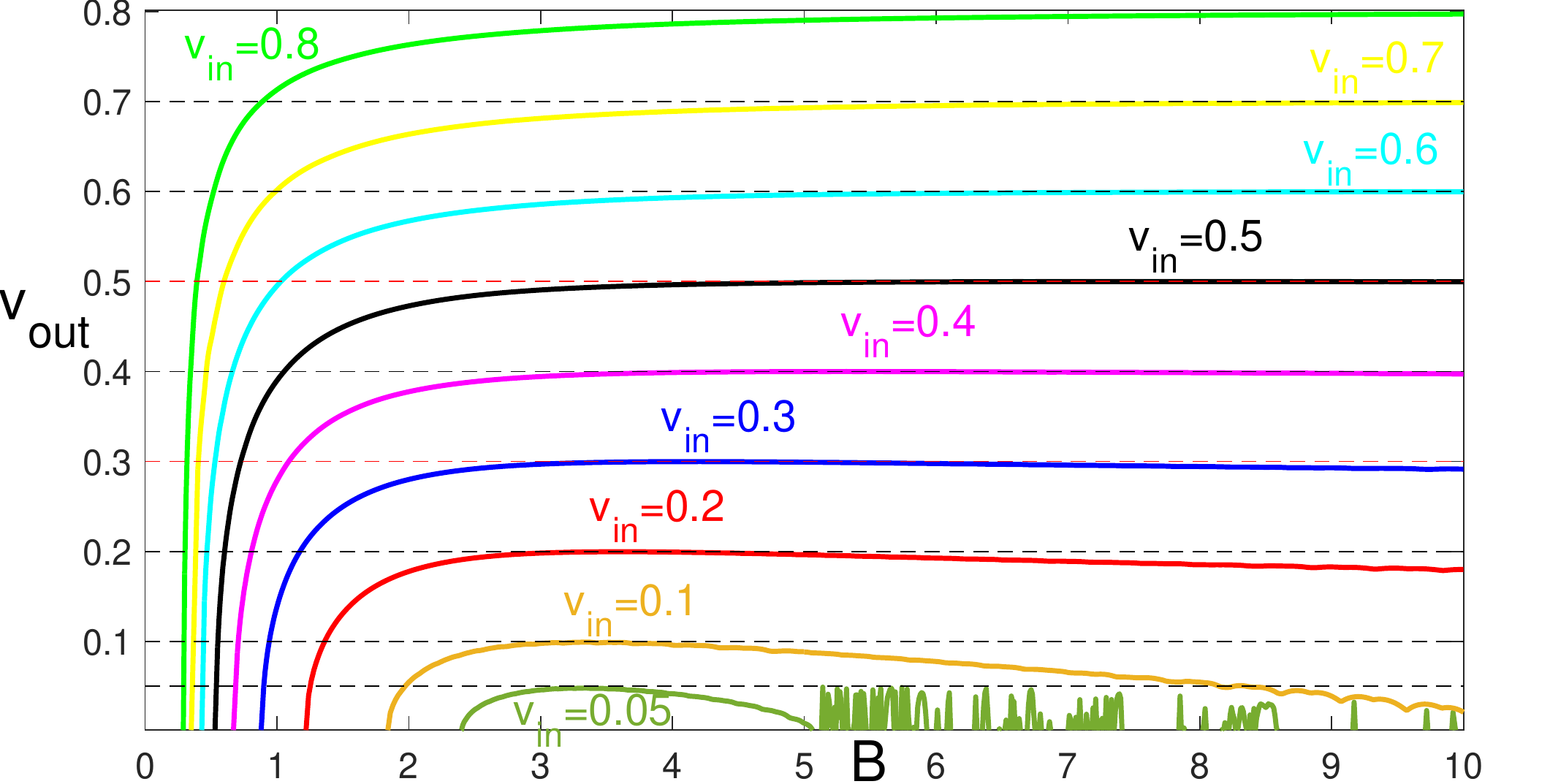}
	\caption{The output velocity versus $B$ for some fixed initial velocities. It should be noted that the curves are not plotted for the whole region of B from zero to $10$ and are considered from their last zero in the region of $B<3.3$ to $B=10$.} \label{voutvsBvinfixed}
\end{figure}

\begin{figure}[htp]	
	\centering	
	\includegraphics[width=150mm]{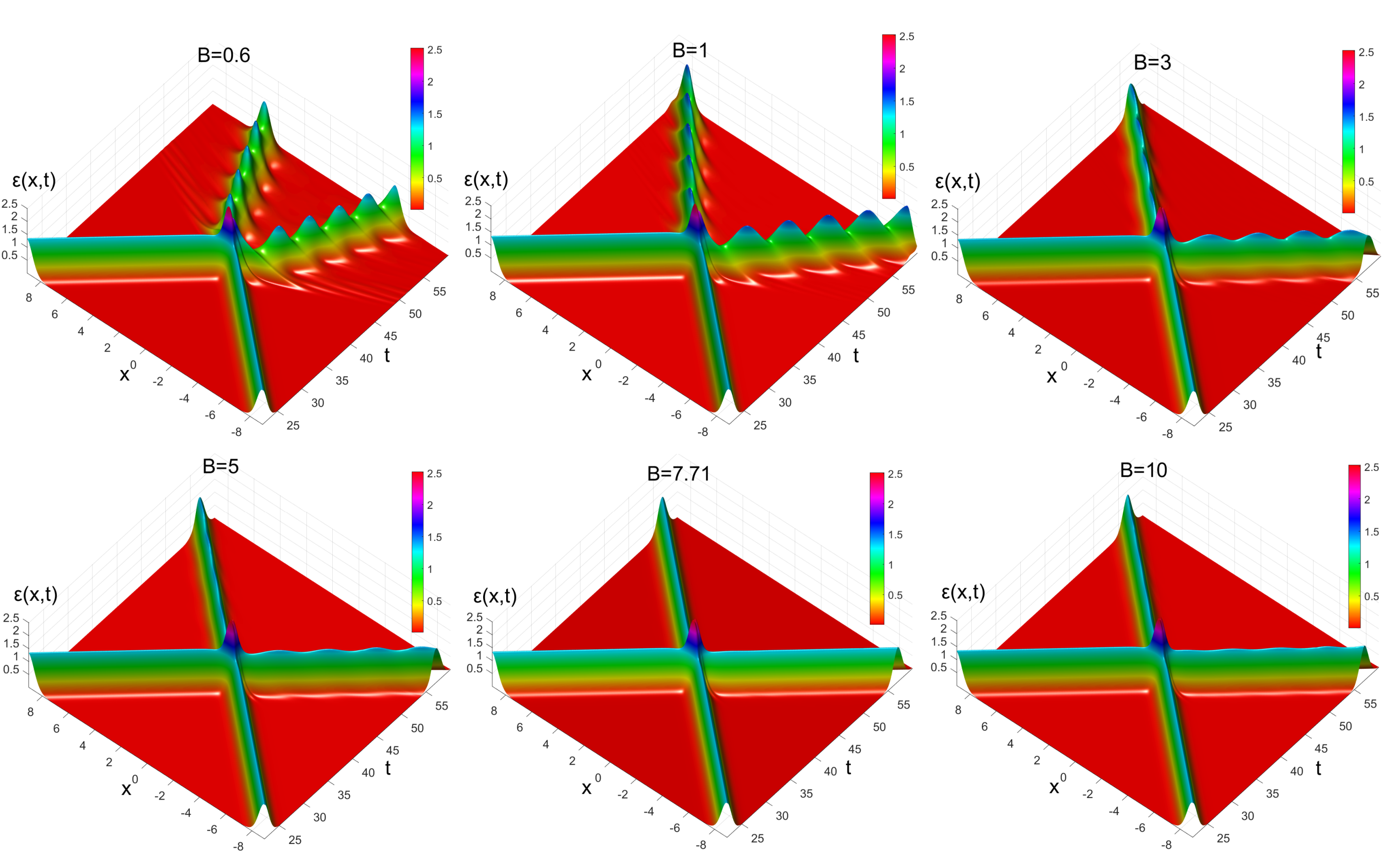}
	\caption{The energy density of kink-antikink collisions for several systems at a fixed initial speed ($v_{in}=0.5$).}	\label{KKvin5}	
\end{figure}

Another diagram which to a large extent reveals the important role of parameter $B$ in
the collisions is the output velocity as a function of $B$ for a constant initial velocity. As shown
in Fig.~\ref{voutvsBvinfixed}, for each constant initial velocity ($v_{in}\geq0.1$ in this study), the corresponding $v_{out}-B$ curve has a maximum at which the output velocity ($v_{out}$) is approximately equal to the initial
velocity ($v_{in}$).
In other words, for these maxima, the behavior of the kink-antikink collisions
is very similar to that of solitons, i.e. their velocities and profiles are restored without any
remarkable change after the collisions. For example, for the case $v_{in}=0.5$, the maximum
occurs at $B=7.71$ and then the kink and antikink have the lowest amplitude oscillations
after the collision (see Fig.~\ref{KKvin5}). These maxima occur at higher $B$s for higher initial velocities.


In general, after each collision, if the kink and anti-kink are scattered from each other, the initial kinetic energy is divided into three types: the kinetic energy of the output kink (anti-kink), the energy related to the internal vibrations, and the energy of low-amplitude fluctuations (radiations) that move at about the speed of light.
The fractions of the radiation and internal vibration energies to the initial kinetic energy (IKE) versus $B$ at different initial velocities result in some minima (see Fig.~\ref{Ein}) that are somewhat close to the maxima in Fig.~\ref{voutvsBvinfixed} (see Table~\ref{maxmin}).
In fact, we expect the share of the radiation and internal vibration energies to be almost minimized when the shape and initial velocity of the kink (anti-kink) are significantly restored. For each initial velocity, an interval (i.e. the intervals between the vertical lines in Fig.~\ref{Ein}-(a)) was found in which the internal energy fraction was less than $10^{-6}$. In these intervals, the midpoints between the parallel lines were considered as the minima.



\begin{table}
	\begin{center}
		\caption{The maximum and minimum points of the curves in Figs.~\ref{voutvsBvinfixed} and \ref{Ein}. }\label{maxmin}
\begin{tabular}{SSSSSSSS} \toprule
	{Initial velocity} & {Color} & {\quad  Maxima  in Fig.~\ref{voutvsBvinfixed}} & {\quad Minima   in Fig.~\ref{Ein}-(a) and (b)}   \\ \midrule
	0.1  &{Brown}& {B=3.34} & {B=3.27,\quad\quad B=3.61}   \\
	0.2  &{Red}& {B=3.64} & {B=3.57,\quad\quad B=4.11}   \\
	0.3  &{Blue}& {B=4.21}  & {B=4.12,\quad\quad B=4.79}  \\
	0.4  &{Purple}& {B=5.32} & {B=5.32,\quad\quad B=5.96}  \\
	0.5  &{Black}& {B=7.71}  & {B=7.64,\quad\quad B=7.26}  \\ \midrule
\end{tabular}
\end{center}
\end{table}


\begin{figure}[ht!]
	\centering
		\includegraphics[width=140mm]{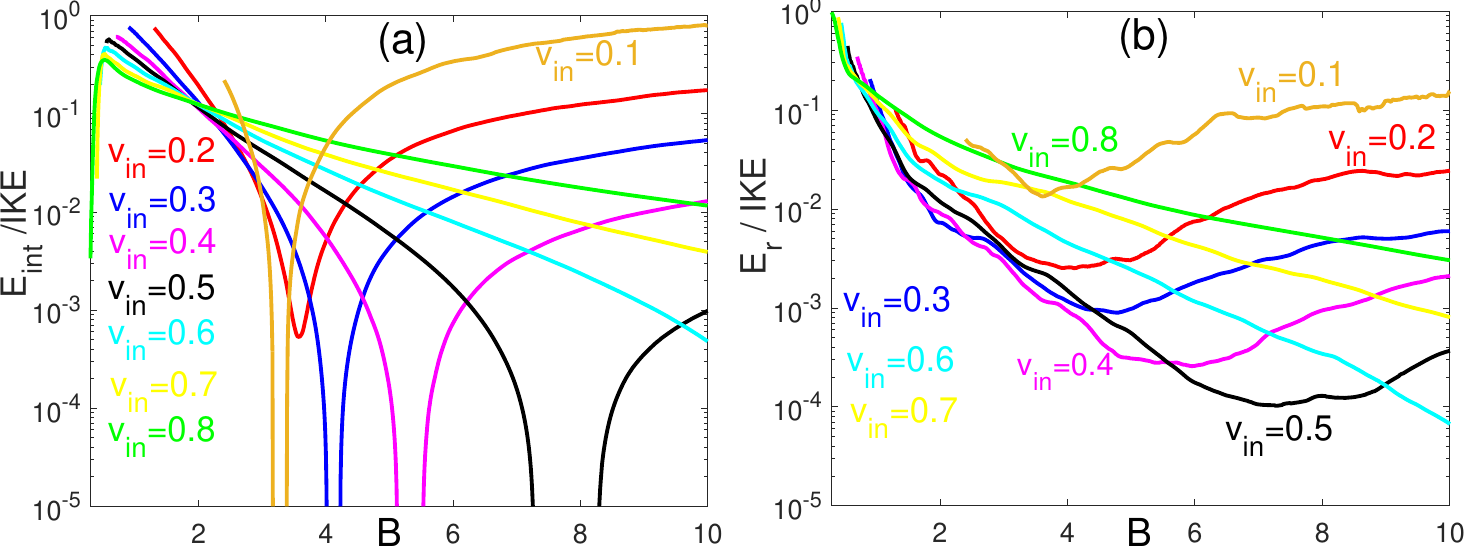}
	\caption{(a) shows the fraction of internal vibration energy to the IKE and (b) illustrates the fraction of radiation energy to the IKE versus $B$ at different initial velocities.} \label{Ein}
\end{figure}

\subsection{The extreme values as a function of $B$}

\begin{figure}[ht!]
	\includegraphics[width=150mm]{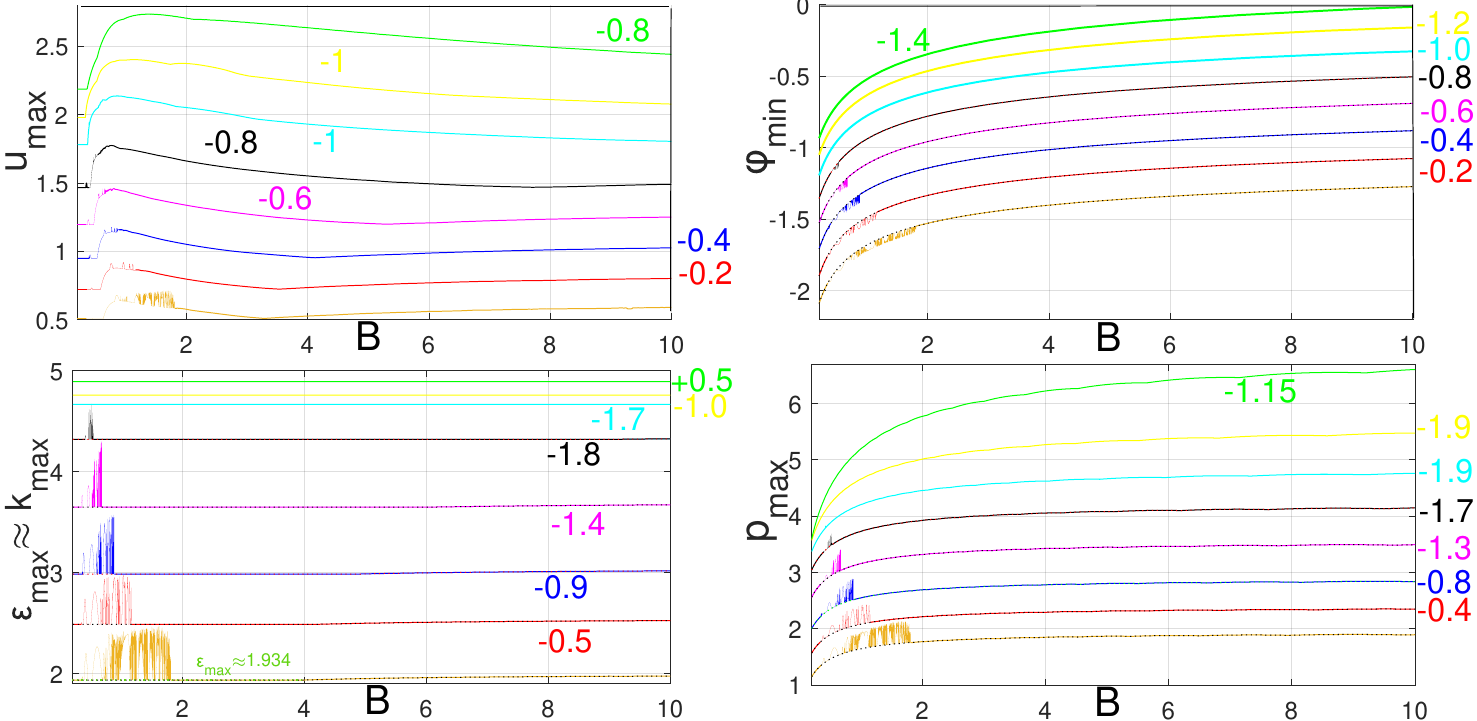}
	\caption{The extreme values as a function of $B$ for different quantities at the center of
		mass point for different initial velocities. The brown, red, blue, purple, black, cyan, yellow,
		and green curves are related to the initial velocities ($v_{in}=0.1, 0.2, \cdots, 0.8$), respectively. To prevent the curves from overlapping and to make them clearly distinct, we had to shift some of them by certain values. To obtain the correct curves, the numbers written on each curve must be added.} \label{extrr}
\end{figure}

For many applications, it is important to know how the maximal and minimal (extreme)
values of different quantities vary versus initial velocities at the center of mass point in
kink-antikink collisions. This subject has been investigated especially for the SG, DSG, $\varphi^4$  and $\varphi^6$ systems \cite{phi412,HighEnergy,HighEnergy2,HighEnergy3,HighEnergy4,HighEnergy5}.
Meanwhile, determining the extreme values of various quantities such as $p$, $k$, $u$, and $\varphi$ provides clearer information about the details of the interactions (collisions). In this regard, we prepare a numerical diagram of the extreme values as a function of $B$ at some initial velocities for different quantities (see Fig.~\ref{extrr}).
For high velocities, all curves change smoothly. However, as the initial velocity decreases, an interval with chaotic behavior appears. The width of this interval increases and becomes more chaotic as the initial velocity decreases.
For such chaotic intervals, there is a quasi-fractal structure in the $\varepsilon_{max}-B$ and $p_{max}-B$
diagrams. For example, if we zoom in on the chaotic interval of the $\varepsilon_{max}-B$  diagram for
the case $v_{in}=0.4$ (the green box in Fig.~\ref{dfr}-(a)), Fig.~\ref{dfr}-(b) is obtained.
 Now zooming
in on another chaotic interval in Fig.~\ref{dfr}-(b) (the red box) leads to a similar diagram
(Fig.~\ref{dfr}-(c)). This process continues in the same way and a finite number of successive zooms results in diagrams almost identical to the prototype (the green box). Moreover,
it is numerically observed that the $\varepsilon_{max}-B$   and  $k_{max}-B$ diagrams are approximately
the same, i.e. $\varepsilon_{max}\approx k_{max}$.
Accordingly, since the maximum value of energy density is expected  to occur  at a space-time $(\tilde{x}=0, \tilde{t})$ with field $\tilde{\varphi}=\varphi(0,\tilde{t})$, according to relation  $\varepsilon=k+u+p=\frac{1}{2}\dot{\varphi}^{2}+\frac{1}{2}\varphi'^{2} +V(\varphi)$, we obtain   $\left.\varphi' \right|_{(0,\tilde{t})}\approx  0$ and $V(\tilde{\varphi})\approx 0 $, i.e. $\tilde{\varphi}\approx -1$.

Another interesting point to consider in Fig.~\ref{extrr} is that the maximum values of $\varepsilon_{max}\approx k_{max}$  remain constant with a good approximation for a long interval of parameter $B$. This result is the same regardless of whether they scatter or stick together and form a bion state. For example, for the case $v_{in}=0.1$, this constant would be  $\varepsilon_{max}\approx 1.934$
in the range of $0.2<B<3.91$ (except for the chaotic interval).
Even in the chaotic intervals, there are smaller sub-intervals for which this situation is valid. For high initial velocities (typically for $v_{in}\geq0.6$ according to Fig.~\ref{dfr}), this range practically covers the entire range (from $0.2$ to $10$) of parameter $B$. Unlike $\varepsilon_{max}$ and $k_{max}$, the maximum values of the potential and gradient energy density (i.e. $p_{max}$ and $u_{max}$) change significantly versus $B$.

\begin{figure}[ht!]
	\centering
	\includegraphics[width=130mm]{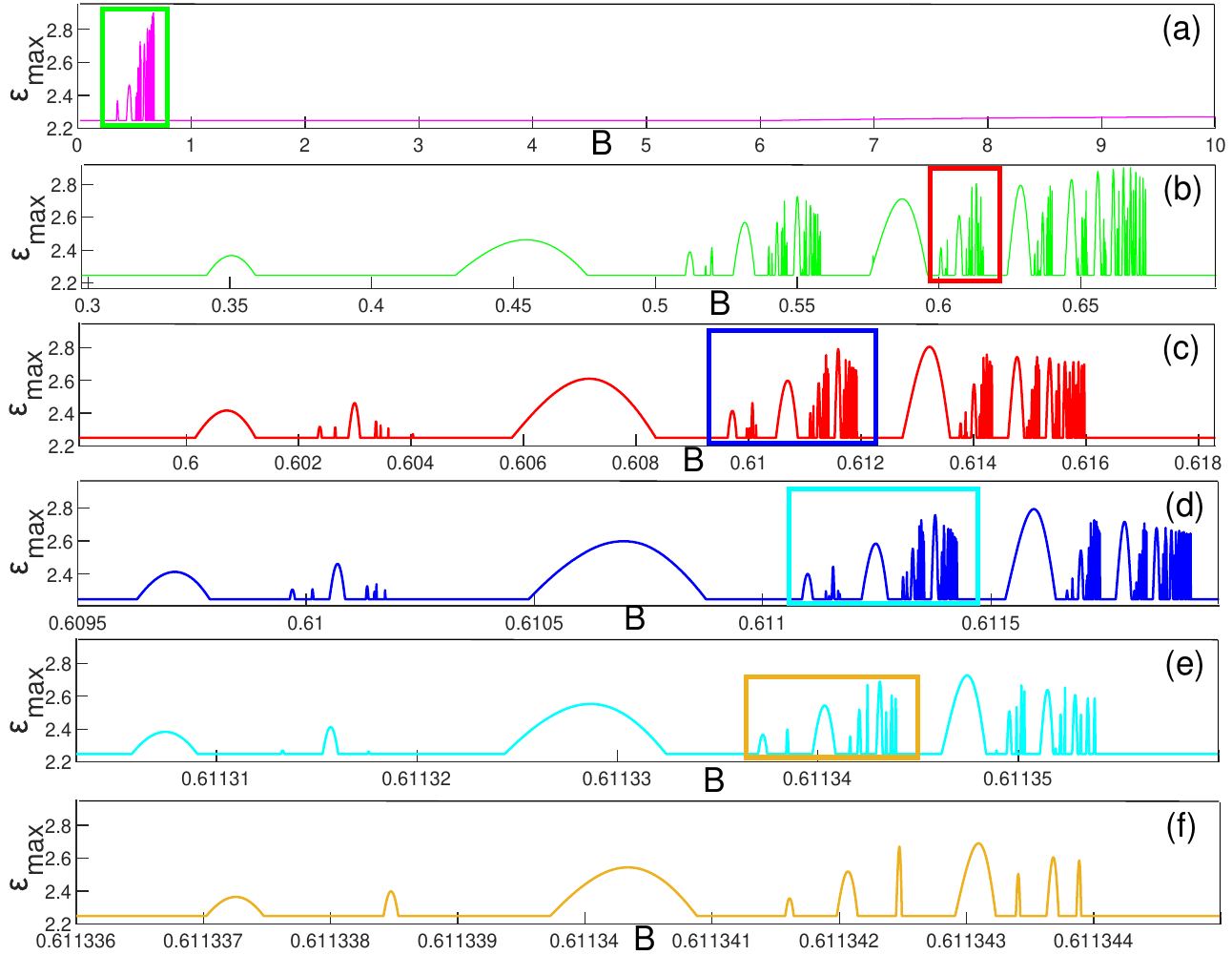}
	\caption{A quasi-fractal structure can be seen on the chaotic region in the $\varepsilon_{max}-B$ diagrams. These plots are obtained for the initial velocity $v_{in}=0.5$ and plot (a) shows the same purple curve in Fig.~\ref{extrr}. By successively zooming in on the rectangular areas, a situation similar to the prototype is obtained only in a finite number of successive zooms.} \label{dfr}
\end{figure}

\begin{figure}[ht!]
	\centering
	\includegraphics[width=145mm]{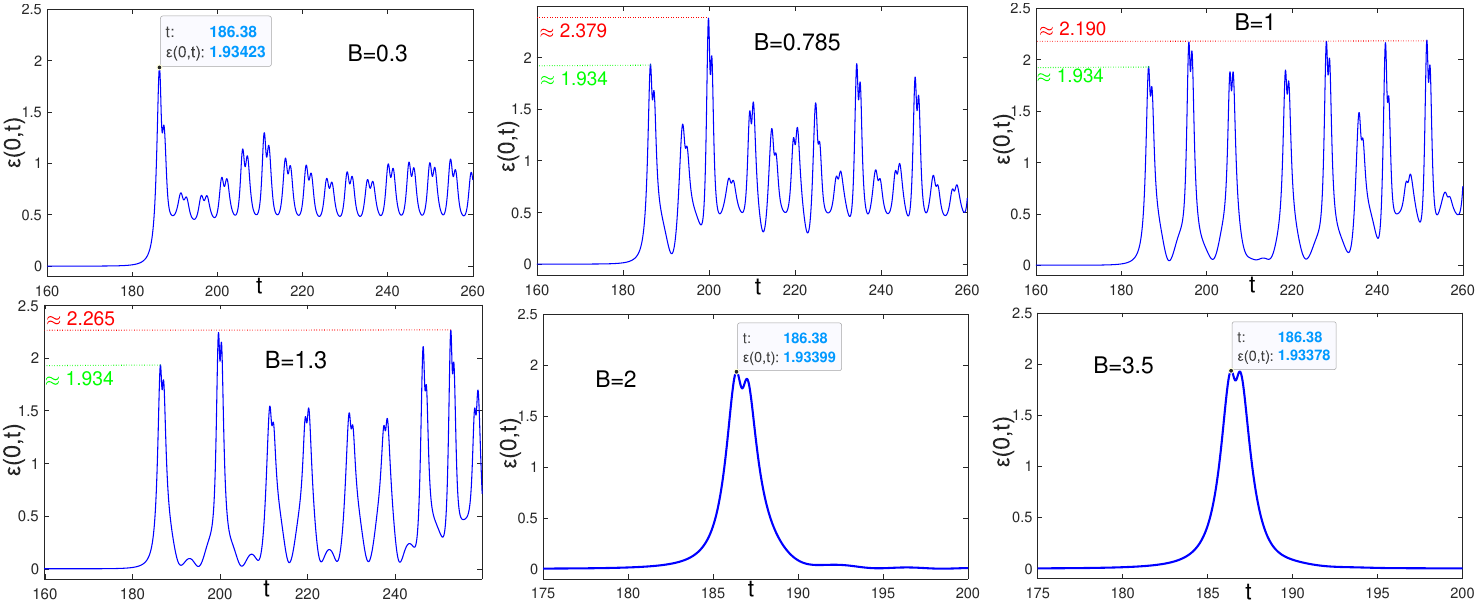}
	\caption{The variations of energy density at the center of mass for a fixed initial velocity ($v_{in}=0.1$). For different systems in the range of $0.23<B<3.91$, the height of the first peak always remains approximately $1.934$.} \label{hj}
\end{figure}

Comparing the energy density diagrams at the center of mass versus time for
different systems at a fixed initial velocity reveals more details about the non-dependence of
the maximum values of energy density on $B$ in a special range.
In this regard, Fig.~\ref{hj} shows that the first maxima of the energy density always remain constant  for $v_{in}=0.1$. Thus, if we only consider the first peaks, a green dotted constant line ($\varepsilon_{max}\approx 1.934$) is obtained in the range $0.2<B<3.91$ (Fig.~\ref{extrr}). In the same way, such an investigation can be repeated for the $p(0,t)$, $k(0,t)$, and $\varphi(0,t)$ diagrams.
In fact, the chaotic regions indicate systems for which the formation of a bion state can be expected in a kink-antikink collision at a fixed initial velocity. In the formation process of a bionic state, several successive peaks are always observed in the $p(0,t)$, $k(0,t)$, and $\varepsilon(0,t)$ representations whose heights may not necessarily be less than that of the first peak (related to the first collision).
If we consider only the first peak to plot the maximal values of $p$, $k$ and $\varepsilon$ against $B$, we will obtain dotted curves in Fig.~\ref{extrr} in which the chaotic regions will entirely disappear.
In other words, the appearance of chaotic regions is due to the peaks (other than the first peak) with larger heights.
It should be noted that in the $\varphi(0,t)$ representation of a bion state, the depth of the successive valleys must be considered to obtain the minimal values of $\varphi$ versus $B$.
Furthermore, since the maximum values of $u(0,t)$ normally occur slightly after the first collision, the $u_{max}-B$ curves behave differently.





\section{Summary  and conclusion}


In this paper, a new version of the famous $\varphi^4$ system (called the $B\varphi^4$ system) was introduced in which the potential was multiplied by a positive parameter $B$ in the range of $\varphi<-1$ and $\varphi>1$  but was considered unchanged at $-1 \leq\varphi \leq1$.
The single kink (anti-kink) solution and its non-interactive properties are independent of parameter $B$ and are the same as those of the typical $\varphi^4$ ($B=1$) system.
However, when a kink interacts with an anti-kink, parameter $B$ dramatically affects the interaction properties. Parameter $B$ can be any arbitrary positive constant. Nevertheless, in this paper, we restricted ourselves to the study of the cases $0.2\leq B\leq 10$. This paper attempted to present a comparative study of the collision details in terms of parameter $B$.


Since the $B\varphi^4$ system is not integrable and multi-kink-antikink solutions are not available
analytically, a numerical scheme based on the method of lines and the Runge-Kutta method
with appropriate accuracy was used to simulate the collisions. In the next step, instead
of preparing $v_{out}-v_{in}$ diagrams, normalized color bar representations (Fig.~\ref{norm}) were
used to obtain and classify the different collisions and escape  windows in numerous systems.
This allowed us to find an approximate curve of the critical velocity variations versus parameter $B$ (Fig.~\ref{vcrvsB}). Accordingly, if parameter $B$ is approximately less than $0.2$, the critical velocity is roughly the speed of light. In other words, if the shape of the potential is inclined to horizontal lines in the regions $\varphi<-1$ and $\varphi>1$, the critical velocity tends to $1$.
On the other hand, it can be expected that the critical velocity tends to zero as $B$ increases. In other words, the closer the potential curves in the regions $\varphi<-1$ and $\varphi>1$ are to verticality, the lower the expected critical velocity. In fact, we expect that larger values of $B$ correspond to stronger repulsive interactions.
This expectation is satisfied up to $B=3.3$ and we see a smooth decrease in the critical velocity curve versus $B$. However, for $B>3.3$, contrary to the expectations, this behavior changes and the $v_{cr}-B$ curve increases slowly.

For the systems $B<1$, the appearance of a quasi-fractal structure in the normalized
$v_{out}-v_{in}$ diagrams is numerically confirmed, i.e. near the boundary of each ${\cal N}$-bounce
window, the distribution of other (${\cal N}+i$)-bounce windows (if any) forms a structure
similar to the prototype.
For the regime $1<B <3.3$, as $B$ increases, the existence of a fractal order is further disturbed due to the fuzzing and shuffling of the escape  windows.
In other words, the prototype-like structure only appears for a number of consecutive zooming in on the edges of the escape  windows.
The further we go from $B=1$, the lower this number. For systems close to $B=3.3$, a completely chaotic behavior is observed in such a way that the colored narrow bars (related to the collisions with different bounces) are thoroughly fuzzed and shuffled.
This chaotic behavior decreases when $B$ becomes larger than $3.3$. The highest order is seen around $B=6$. However, it is not enough to see a fractal structure. Thereafter, the chaotic behavior slowly increases as $B$ increases. In general, no quasi-fractal structure was seen in the regime $3.3<B<10$.


The fuzzing of various escape  windows dissolves their clear boundaries. Particularly, it was observed that different two-bounce windows were clearly intertwined. The peak of this intertwining occurs for systems close to $B=3.3$. In general, this also applies to other  ${\cal N}$-bounce windows.  
Thus, a  two-bounce escape  window  should be defined as a collection of several separate intervals (classified based on having the same number of cycle oscillations).   
Two-bounce escape  windows can be labeled with incremental integers from left to right. The number of cycle oscillations of the first two-bounce window ($\bar{n}$) is not constant and is a function of $B$. As a rule, except for the systems close to $B=3.3$, the number of cycle oscillations of the subsequent two-bounce windows will increase one by one in order.
In this regard, known phenomenological relations (\ref{bn2}) and (\ref{dfv}) were studied and a good agreement between the numerical and theoretical results was obtained (see Figs.~\ref{TC} and \ref{curves}). These results confirmed that the resonance frequency ($\omega_{s}$)  and the kink vibrational mode frequency ($\omega_{o}$) were approximately the same and independent of parameter $B$.

The emergence of a soliton behavior in collisions was another interesting point that was
investigated in this article. It was observed that for a given initial velocity, if the final velocities are examined versus parameter $B$, an interval for parameter $B$ can be specified in which the kink-anti-kink pair leaves the collision site at approximately the same initial velocity without a significant deformation.
To have a criterion for the amount of kink deformation after the collisions, the fraction of internal vibrational energy to the initial kinetic energy ($E_{int}/IKE$) versus $B$ can be considered. In addition, the same was done for the amount of radiation energy released in each kink-antikink collision (i.e. $E_{r}/IKE$). Therefore, we expected the minima in the $E_{int}/IKE-B$ and $E_{r}/IKE-B$ diagrams to correspond to the maximum in the $v_{out}-B$ diagram. This was the case in the current study.

Studying the extreme values of different quantities at the center of mass point of the
kink-antikink collisions versus parameter $B$ reveals more details about the role of parameter $B$ in the interactions. Accordingly, the curves related to the maximum values
of energy density and its different parts (potential, gradient, and kinetic energy density),
as well as the minimal values of the field ($\varphi$) were plotted versus $B$ at certain initial
velocities.
Interestingly, the curves of $\varepsilon_{max}$  and  $k_{max}$  were almost the same and remained almost constant for some intervals of $B$. A chaotic region was observed in each of these diagrams at low initial velocities. The width of this chaotic region decreased as the initial velocity increased. It was shown that there was a quasi-fractal structure for  $\varepsilon_{max}$  and  $p_{max}$ in such regions.
The chaotic region specified the systems in which forming a bion state can be expected for a fixed initial speed. A bion state is an oscillating localized field consisting of a series of peaks and valleys.
Except for the $u_{max}-B$ diagrams, the presence of chaotic regions in  $\varepsilon_{max}-B$, $k_{max}-B$, and $u_{max}-B$ diagrams is related to the peaks (with greater heights) other than the first peak. It is noteworthy that if we consider only the first peak (related to the first collision), the chaotic region does not appear.


Examining the mass numerical results obtained in this article can be a start for future studies.
There are many questions regarding the obtained numerical results which can be considered as open problems for future works. For instance, why is there  a minimum  in the $v_{cr}-B$ curve (Fig.~\ref{vcrvsB})? Does  this curve tend  to a certain value for larger values of $B$ or something else? 
Why does the resonance frequency  remain almost constant  according to Figs.~\ref{TC} and \ref{curves}? What other  factors should  be involved in the occurrence  of the escape-window regions in addition to  the resonance energy exchange phenomenon? Why do the maxima of the first peaks remain constant versus $B$ in the energy density representation  according to Fig.~\ref{hj}? Using  a CCM based on the shape modes or  Derrick modes may help to get an insight as their results depend on  parameter $B$ \cite{phi47,phi4cc,phi4cc2}. Especially in \cite{phi4cc2}, the researchers used a relativistic approach that could predict the critical velocity and overall fractal structure, which showed good agreement with the theory.  This may help to explain the observed dependence of the escape-window region and $v_{cr}$ on $B$ in the current study. 
Also, for the cases where the initial velocities are very small, another phenomenon--the spectral wall phenomenon \cite{Adam1,Adam2}--may be relevant.  Hence, it may be considered in relation to why a chaotic structure appears for systems close to $B=3.3$. It is related to the crossing of a shape mode to the free modes during an evolution (e.~g., the kink-antikink collision) and acts as a barrier for solitons. 
Overall, we believe that the current study opens up a wide perspective for further research on other kink-bearing systems such as SG, $\varphi^6$, and $\varphi^8$ ones. The special deformations introduced in this article for some physical systems, such as the propagation of solitary electromagnetic waves in graphene superlattices, can be interpreted as different interactions for an undeformed (kink and antikink) entity which deserve further investigation.



\section*{Acknowledgement}

The authors  wish to express their appreciation to the Persian Gulf University Research Council for their constant support.

\end{document}